\documentclass[]{aa}
\usepackage{graphicx}
\usepackage{txfonts}
\def\kms{km~s$^{-1}$}

\def\D{{\cal D}}
\def\L{{\cal L}}
\def\O{{\cal O}}
\def\d{{\rm d}}

\def\R{{\cal R}}

\def\t{\vec{\theta}}

\def\x{\vec{x}}

\def\z{\vec{z}}

\def\sM{\sigma_{\rm M}}
\def\sU{\sigma_{\rm U}}
\def\sU'{\sigma_{\rm U'}}
\def\sV{\sigma_{\rm V}}
\def\sV'{\sigma_{\rm V'}}
\def\sW{\sigma_{\rm W}}

\def\ML{{\rm ML}}

\begin{document}
\titlerunning{Local kinematics of K and M giants}
  \title{Local kinematics of K and M giants from CORAVEL/Hipparcos/Tycho-2
  data\thanks{Based on observations performed at the Swiss 1m-telescope at
  OHP, France, and on data from the ESA Hipparcos astrometry satellite}}  
   
  \subtitle{Revisiting the concept of superclusters}

   \author{B. Famaey
          \inst{1}\fnmsep\thanks{Research Assistant, F.R.I.A.}
          \and
          A. Jorissen
          \inst{1}\fnmsep\thanks{Senior Research Associate, F.N.R.S.}
          \and
          X. Luri
          \inst{2}
          \and
          M. Mayor
          \inst{3}
          \and
          S. Udry
          \inst{3}
          \and 
          H. Dejonghe
          \inst{4}
          \and
          C. Turon
          \inst{5}
          }

  \offprints{B. Famaey}

   \institute{Institut d'Astronomie et d'Astrophysique, Universit\'e
Libre de
  Bruxelles, CP 226, Boulevard du Triomphe, B-1050 Bruxelles, Belgium.\\
              \email{bfamaey@astro.ulb.ac.be; ajorisse@astro.ulb.ac.be}
         \and
             Departament d'Astronomia i Meteorologia, Universitat de
	     Barcelona, Avda. Diagonal 647, E-08028 Barcelona, Spain
         \and
             Observatoire de Gen\`eve, Chemin des Maillettes 51, CH-1290
	     Sauverny, Switzerland
         \and
             Sterrenkundig Observatorium, Universiteit Gent, Krijgslaan 281, 
             B-9000 Gent, Belgium  
         \and
             Observatoire de Paris, section de Meudon, GEPI/CNRS UMR 8111,
	     F-92195 Meudon CEDEX, France
             }

   \date{Received ...; accepted ...}

   \abstract{
The availability of the Hipparcos Catalogue triggered many kinematic and
dynamical studies of the solar neighbourhood. Nevertheless, those
studies generally lacked the third component of the space velocities, i.e., the
radial velocities. This work presents the kinematic analysis of
5952 K and 739 M giants in the solar neighbourhood which includes for the
first time radial
velocity data from a large survey performed with the CORAVEL
spectrovelocimeter. It also uses proper motions from the Tycho-2 catalogue,
which are expected to be more accurate than the Hipparcos ones. An
important by-product of this study is the observed fraction of only 5.7\%
of spectroscopic binaries among M giants as compared to 13.7\% for K
giants. After excluding the binaries for which no center-of-mass velocity
could be estimated, 5311 K and 719 M giants remain in the final sample. 

The $UV$-plane constructed from these data for the stars with precise
parallaxes ($\sigma_\pi/\pi \le 20$\%) reveals a rich small-scale
structure, with several clumps corresponding to the Hercules stream, the
Sirius moving group, and the Hyades and Pleiades superclusters.  
A maximum-likelihood method, based on a bayesian approach, has been applied
to the data, in order to make full use of all the available stars (not only
those with precise parallaxes) and to derive the kinematic properties of
these subgroups. Isochrones in the Hertzsprung-Russell diagram
reveal a very wide range of ages for stars belonging to these groups.
These groups are most probably related to the dynamical perturbation by
transient spiral waves (as recently modelled by De Simone et al. 2004) rather than to cluster remnants. A possible explanation for the presence of young group/clusters  in the same
area of the $UV$-plane is that they have been put there by the spiral wave
associated with their formation, while the kinematics of the older
stars of our sample has also been disturbed by the same wave. The emerging
picture is thus one of {\it dynamical streams} pervading the solar neighbourhood and
travelling in the Galaxy with similar space velocities. The term {\it dynamical stream} is more appropriate than the traditional term {\it
  supercluster} since it
involves stars of different ages, not born at the same place nor at the
same time. The position of those streams in the $UV$-plane is responsible
for the vertex deviation of $16.2^\circ \pm 5.6^\circ$ for the whole
sample. Our study suggests that the vertex deviation for younger
populations could in fact have the same {\it dynamical} origin.  

The underlying velocity ellipsoid, extracted by the maximum-likelihood
method after removal of the streams, is not centered on the value commonly
accepted for the radial antisolar motion: it is centered on $\langle U
\rangle = -2.78\pm1.07$~\kms. However,
the full data set (including the various streams) {\it does} yield the
usual value for the radial solar motion, when properly accounting for the
biases inherent to this kind of analysis (namely, $\langle U \rangle =
-10.25\pm0.15$~\kms). This discrepancy clearly raises the essential question
of how to derive the solar motion in the presence of dynamical
perturbations altering the kinematics of the solar neighbourhood: does
there exist in the solar neighbourhood a subset of stars having no net
radial motion which can be used as a reference against which to measure the
solar motion? 
  
}
\maketitle
%
%________________________________________________________________

\section{Introduction}

The kinematics of stars in the solar neighbourhood, studied from proper
motions and radial velocities, has long been acknowledged to be an
essential element for the understanding of the structure and evolution of
the Galaxy as a whole\ (e.g. Schwarzschild 1907, Lindblad 1925, Oort
1927). The story of the efforts to obtain the necessary data is standard
textbook fare, and nicely illustrates how theoretical progress and data
acquisition have to go hand in hand if one wants to gain insight in the
true nature of our Galaxy. Clearly, another chapter in this story has begun
with the Hipparcos satellite mission\ (ESA
1997), that provided accurate parallaxes and proper motions for
a large number of  stars (about 118\ts 000): with
accurate parallaxes available, it was no longer necessary to
resort to photometric distances.
The Hipparcos data enabled several kinematic studies
of the solar neighbourhood, but those studies generally lacked
radial velocity data. A first preview of how Hipparcos data could 
improve our knowledge of stellar motions in the Galaxy 
was given by Kovalevsky\ (1998); then, Dehnen \&
Binney\ (1998) used Hipparcos proper motions to derive some fundamental
kinematic parameters of the solar neighbourhood.
They did not use radial velocities from the literature
because, at that time, radial velocities had been measured preferentially
for high proper motions stars and their use would have introduced a
kinematic bias. Hipparcos data were also used by Chereul et al.\ (1998) to
derive the small scale structure of the velocity distribution of early-type
stars in the solar neighbourhood. 

The situation has dramatically improved thanks to the efforts of a
large European consortium to obtain radial velocities of Hipparcos stars 
later than about F5 (Udry et al. 1997). The sample includes Hipparcos 
`survey' stars (flag S in field H68 of the Hipparcos Catalogue), and 
stars from other specific programmes.    
This unique database, comprising about 45\ts 000 stars measured with the 
CORAVEL spectrovelocimeter (Baranne et al. 1979) at a typical accuracy of 
0.3~\kms, combines a  high precision and the absence of kinematic bias. It
thus represents an unprecedented data set to test the results obtained by
previous kinematic studies based solely on Hipparcos data. Another
feature of the present study is the use of Tycho-2 proper motions,
which combine Hipparcos positions with positions from much older
catalogues (H{\o}g et al. 2000). They represent a subsantial
improvement over the Hipparcos proper motions themselves, which are based
on very accurate positions, but extending over a limited 3-year time
span. Results from the Geneva-Copenhagen survey for about 14\ts 000 F and G
dwarfs present in the CORAVEL database have been recently published by
Nordstr\"om et al.\ (2004). The present paper, on the other hand,
concentrates on the K and M giants from the CORAVEL database.

According to the classical theory of moving groups,
early-type stars are not good probes to determine global characteristics of 
the Galaxy because these stars are young and still carry the kinematic
signature of their place of birth. As a 
consequence, their distribution in velocity space is clumpy\
(e.g. de Bruijne et al. 1997, Figueras et
al. 1997, de Zeeuw et al. 1999). These inhomogeneities can be spatially confined groups of young
stars (OB associations in the Gould's belt, young clusters) but can also be spatially
extended groups. There is, however, some confusion in the literature about the related
terminology. Eggen (1994) defines a 'supercluster' as a group of stars
gravitationally unbound that share the same kinematics and may occupy
extended regions in the Galaxy, and a 'moving group' as the part of
the supercluster that enters the solar neighbourhood and can be
observed all over the sky. Unfortunately, the same term 'moving group' is
sometimes also applied to OB associations (e.g. de Zeeuw et al. 1999). It
has long been known that, in the solar
vicinity, there are several superclusters and moving groups that share the
same space motions as well-known open 
clusters\ (Eggen 1958). The best documented groups (see Montes et
al. 2001 and references therein) are the Hyades
supercluster associated with the Hyades cluster (600 Myr) and the Ursa
Major group (also known as the Sirius supercluster) associated with the
UMa cluster of stars (300 Myr). Another kinematic group called the
Local Association or Pleiades moving group is a reasonably coherent
kinematic group with embedded young clusters and associations like the
Pleiades, $\alpha$ Persei, NGC 2516, IC 2602 and Scorpius-Centaurus, with
ages ranging from about 20 to 150 Myr. Two other young moving groups
are the IC 2391 supercluster (35-55 Myr) and the Castor moving group
(200 Myr). The kinematic properties of all these moving groups and
superclusters are listed by Montes et al. (2001) and Chereul et
al. (1999). However, the recent observation (e.g. Chereul et al. 1998,
1999) that some of those superclusters involve early-type stars spanning a wide
range of ages contradicts the classical hypothesis that supercluster stars
share a common origin:
Chereul et al.\ (1998) propose that the supercluster-like velocity
structure is just a chance juxtaposition of several cluster
remnants. 

Nevertheless, significant clumpiness in velocity space has also
been recently reported for {\it late-type} stars\ (Dehnen 1998), thus
raising the
questions of the age of those late-type stars and of the exact origin of this
clumpiness. For example, Montes et
al.\ (2001) suppose that, if a late-type star belongs to a moving group or
supercluster, it may be considered as a sign that the star is young, an
assumption that will be largely challenged by the results of this
paper. Here, we attack the problem from another point of view by restricting
our study of stellar kinematics to late-type giants (of spectral types K
and M), without any a priori hypothesis on their motion or age.  

The present paper is organized as follows. 
Sect.~2 describes the various data sets used in the present study.
The kinematic analysis of the data and its implications for the dynamics of our
Galaxy are presented in Sect.~3. For summary and conclusions we refer to
Sect.~4. 
  
\section{The data}

\subsection{Selection criteria}

\begin{figure}
   \centering
   \includegraphics[width=8cm]{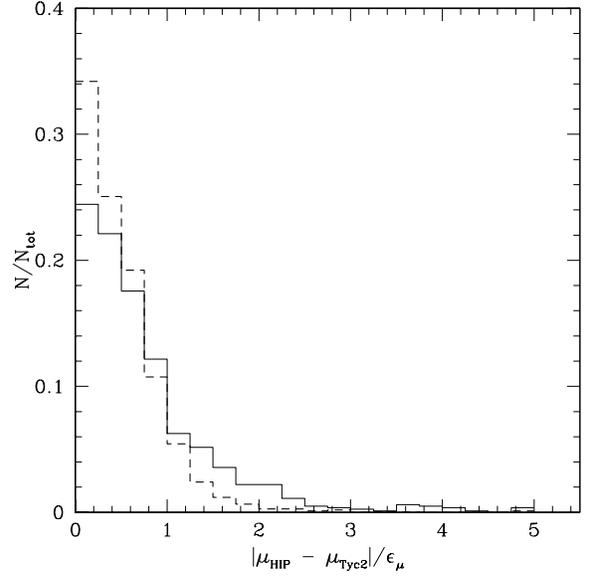}
      \caption{\label{Fig:Tycho2}
Difference between the Hipparcos and Tycho-2 proper-motion moduli, normalized by the root mean square
of the standard errors on the Hipparcos and Tycho-2 proper motions,
    denoted $\epsilon_\mu$. The solid line refers to the 859 spectroscopic binaries (SB)
present in our sample, whereas the dashed line corresponds to the 5832
non-SB stars. Note that the sample of binaries contains more cases where
the Hipparcos and Tycho-2 proper motions differ significantly, in agreement
with the argument of Kaplan \& Makarov\ (2003).
              }
   \end{figure}

Our stellar sample is the intersection of several data sets: (i) in a
first step, all stars with spectral types K and M appearing in field H76
of the Hipparcos Catalogue (ESA 1997) have been selected, and the
Hipparcos parallaxes have been used in the present study;  (ii) Proper
motions were taken from the Tycho-2 catalogue (H{\o}g et al. 2000). These
proper motions are more accurate than the Hipparcos ones.
Fig.~\ref{Fig:Tycho2} compares the proper motions from Tycho-2 and Hipparcos, and reveals that the Tycho-2 
proper motions are sufficiently different from the Hipparcos ones
(especially for binaries) to 
warrant the use of the former in the present study; 
(iii) Radial-velocity data for stars belonging to this
first list (stars with spectral type K and M) have then been retrieved from
the CORAVEL database.  
Only stars from the northern hemisphere ($\delta > 0^\circ$), observed with
the 1-m Swiss telescope at the {\it Haute-Provence Observatory}, have been
considered. As described by Udry et al. (1997), all stars from the
Hipparcos survey (including all stars brighter than $V = 7.3 + 1.1 |\sin b|$
for spectral types later than G5, where $b$ is the galactic latitude; those
stars are flagged `S' in field H68 of the Hipparcos Catalogue) are 
present in the CORAVEL database. Besides the survey, stars from several other 
Hipparcos programmes, dealing 
with galactic studies and requiring radial velocities, were also monitored with
CORAVEL. The CORAVEL database finally contains stars
monitored for other purposes (e.g., binarity or rotation), but they
represent a small fraction of the Hipparcos survey stars.

\begin{figure}
   \centering
   \includegraphics[width=8cm]{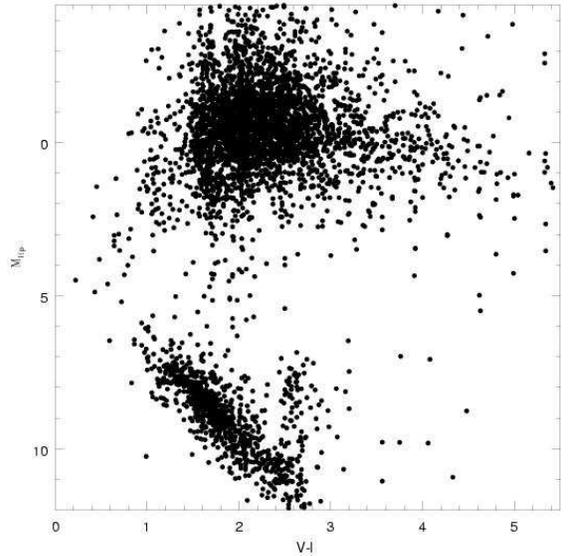}
      \caption{
\label{Fig:DHRM}
Crude Hertzsprung-Russell diagram for the Hipparcos M stars with positive
parallaxes.
              }
   \end{figure}

\begin{figure}
   \centering
   \includegraphics[width=8cm]{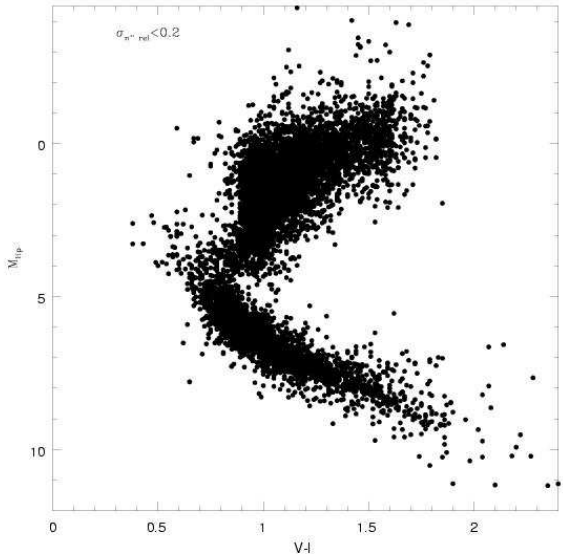}
      \caption{
\label{Fig:DHRK}
Crude Hertzsprung-Russell diagram for the Hipparcos K stars with
a relative error on the parallax less than 20$\%$.
              }
   \end{figure}

The sample resulting from the intersection of these 3 data sets was
further cleaned as follows.
First, a crude Hertzsprung-Russell diagram for all the M-type stars of the
Hipparcos Catalogue has been constructed from distances estimated from a simple inversion of 
the parallax (Fig.~\ref{Fig:DHRM}), and
reveals a clear separation between dwarfs and giants. Clearly,
all M stars with $M_{Hp}<4$ must be giants. For K-type stars with a
relative parallax error less than 20\%, the Hertzsprung-Russell diagram  
(Fig.~\ref{Fig:DHRK}) shows that the giant branch connects to the main
sequence around
$(V-I=0.75, M_{Hp}=4)$. Therefore, in order to select only K and M
giants, only stars with
$M_{Hp}<4$ were kept in the sample. For K stars, the 
stars with  $V-I<0.75$ and
$M_{Hp}>2$ were eliminated to avoid contamination by K dwarfs.
Diagnostics based on the radial-velocity variability and
on the CORAVEL cross-correlation profiles, combined with literature
searches, allowed us to further screen out stars with peculiar spectra,
such as T~Tau stars, Mira variables and S stars. T~Tau stars have been
eliminated because they belong to a specific young population; S stars because
they are a mixture of extrinsic and intrinsic stars, which have different
population characteristics (Van Eck \& Jorissen 2000);  Mira variables
because their center-of-mass radial velocity is difficult to derive from
the optical spectrum, where confusion is introduced by the pulsation
(Alvarez et al. 2001).

The primary sample includes 5952 K giants and 739 M giants (6691 stars).
86\% of those stars are 'survey' stars: our sample is thus complete for the
K and M giants brighter than $V = 7.3 + 1.1 |\sin b|$.  To fix the ideas, for a typical giant star with $M_V=0$, this magnitude threshold translates 
into distances of  290~pc in the galactic plane and 480~pc in the direction of the galactic pole 
(these values are, however, very sensitive upon the adopted absolute magnitude, and become  45~pc    
for a subgiant star with $M_V = +4$ in the galactic plane, and 2900~pc for a supergiant star 
with $M_V= -5$, corresponding to the range of luminosities
present in our sample; see Sect.~\ref{Sect:groups}).
%{\bf This means that
%  {\it all} the existing giants (as defined in this section) of the
%  northern hemisphere are present in
%  the sample out to the distance of 50 pc in the galactic plane and out to
%  85 pc in the direction of the galactic pole.} 
The final step in the
preparation and cleaning of the sample involves the identification of the
spectroscopic binaries, as described in Sect.~\ref{Sect:binaries}.

\subsection{Binaries}
\label{Sect:binaries}

The identification of the binaries, especially those with large
velocity amplitudes, is an important step in the selection process, because
kinematic studies should make use
of the center-of-mass velocity. In order to identify those,
the observing strategy was to obtain at least two radial-velocity
measurements per star, spanning 2 to 3~yr. Monte-Carlo simulations reveal
that, with such a strategy,  binaries are detected with an
efficiency better than 50 percent (Udry et al. 1997). 

\begin{figure}
   \centering
   \includegraphics[width=8cm]{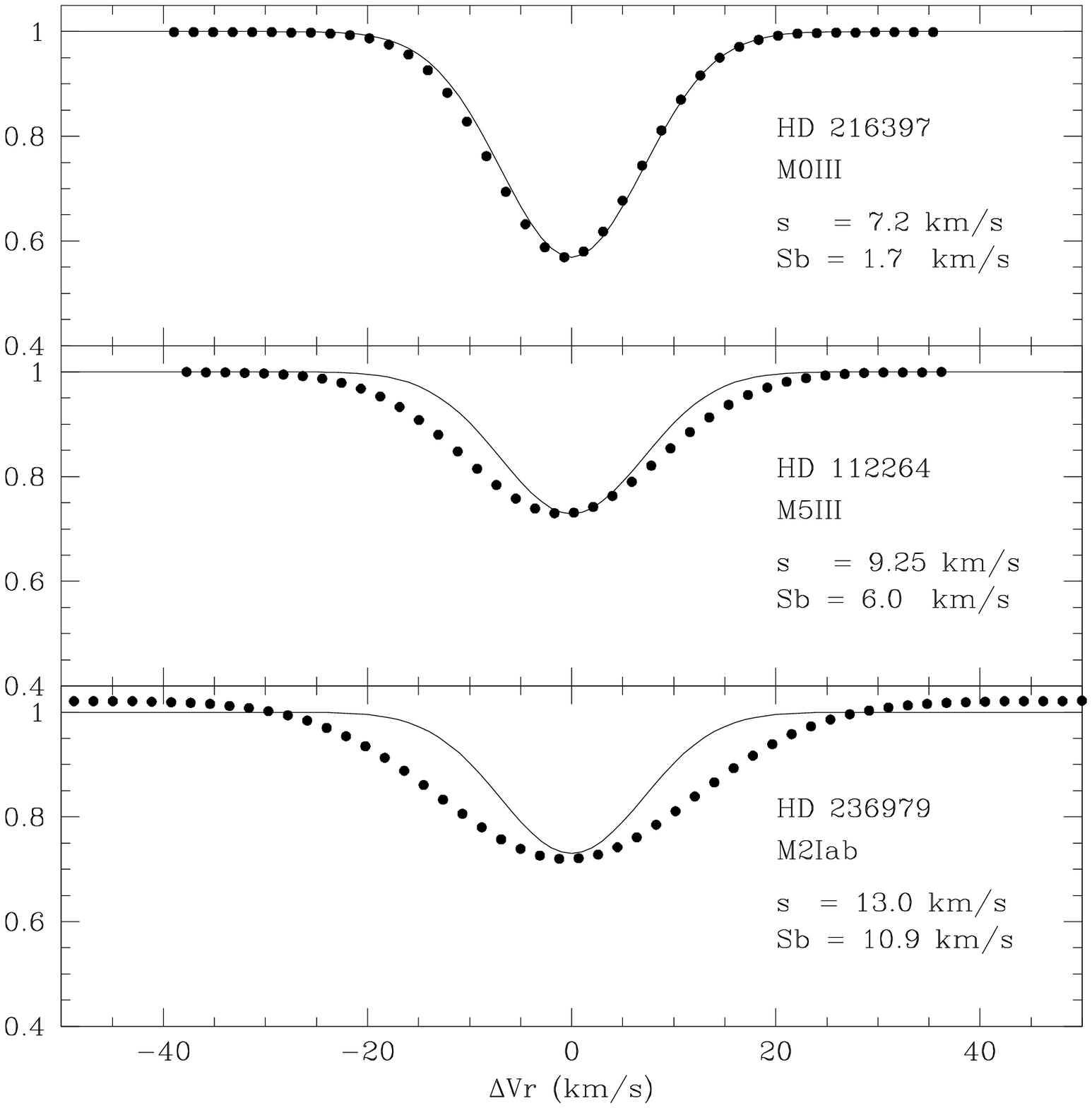}
      \caption{\label{Fig:Sbprofil}
The concept of line-width parameter $Sb = (s^2 - s_0^2)^{1/2}$ 
is illustrated here by comparing
the CORAVEL cross-correlation (smoothed) profiles (black dots) of 3 M
giants or supergiants with the gaussian instrumental profile (of sigma
$s_0 = 7$~\kms; solid line). Each profile corresponds to a single
radial-velocity measurement for the given stars.
The $s$ value listed in the above panels refers
to the sigma parameter of a gaussian fitted to the observed correlation
profile.    
              } 
   \end{figure}

\begin{figure*}
   \centering
   \includegraphics[width=15cm]{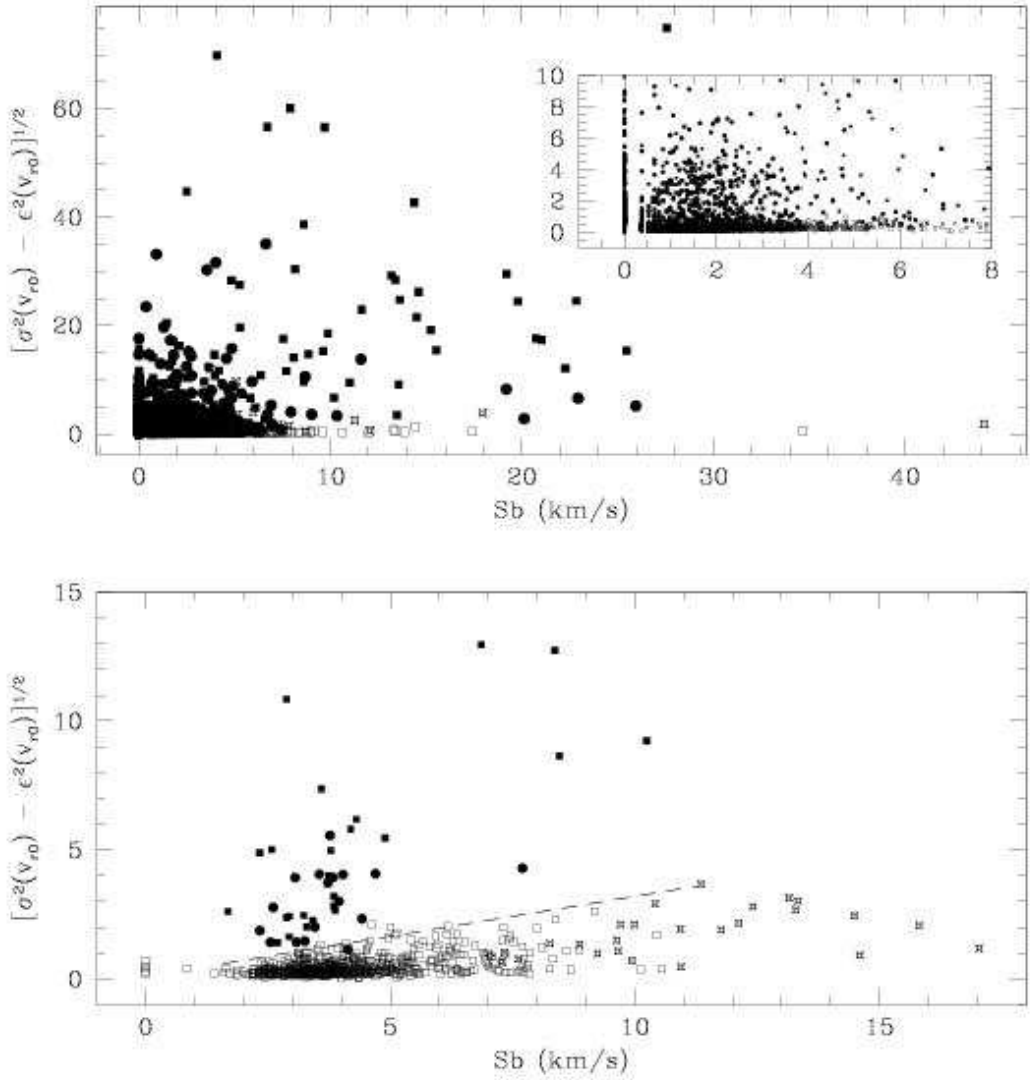}
      \caption{\label{Fig:Sbsigma}
The ($Sb, \sigma_0(v_{r0}))$ diagram (see text) for K  (upper panel, and
zoom inside) and M
giants (lower panel). Star symbols denote supergiant stars, filled symbols
denote 
spectroscopic binaries  with (squares) or without (circles) center-of-mass 
velocity available.
              }
   \end{figure*}

Since late-type giants exhibit intrinsic radial velocity jitter
(Van Eck \& Jorissen 2000), the identification of
binaries requires a specific strategy, which makes use of the  ($Sb,
\sigma_0(v_{r0}))$ diagram (Fig.~\ref{Fig:Sbsigma}), where $v_{r0}$ is the
measured radial velocity. The parameter $Sb$ is a measure of the intrinsic
width of the cross-correlation profile, {\it i.e.,} corrected from the instrumental
width (see Fig.~\ref{Fig:Sbprofil} for more precise definitions). 
The CORAVEL instrumental profile  is obtained from the
observation of the cross-correlation dip of minor planets reflecting the 
sun light, after correction
for the solar rotational velocity and photospheric turbulence (see Benz \&
Mayor 1981 for more details). The
$Sb$ parameter is directly related to the average spectral line width,
which is in turn a function of spectral type and luminosity, the
later-type and more luminous stars having larger $Sb$ values. On the other
hand, the measurement error $\bar{\epsilon}$ has been quadratically subtracted
from the  radial velocity standard  deviation
$\sigma(v_{r0})$ to yield the effective  standard  deviation
$\sigma_0(v_{r0})$.

The identification of binaries among K or M giants follows different steps.
The standard $\chi^2$ variability test, comparing the standard deviation $\sigma(v_{r0})$ of the 
measurements to their
uncertainty $\bar{\epsilon}$, cannot be applied to M giants,  because intrinsic radial-velocity 
variations 
(`jitter') associated with envelope pulsations would flag them as velocity variables 
in almost all cases. Therefore, in a first step,
M giants having 
$\sigma_0(v_{r0}) \ge 1$~\kms  have been monitored with the ELODIE
spectrograph (Baranne et al. 1996) at the {\it Haute-Provence Observatory}
(France) since August 2000 (Jorissen et al. 2004). These supplementary data
points made it possible to distinguish orbital variations from
radial-velocity jitter by a simple visual inspection of the data. 
It turns out that, in the ($Sb, \sigma_0(v_{r0}))$ diagram, all the confirmed
binaries (filled symbols in the lower panel of Fig.~\ref{Fig:Sbsigma})
are located in the upper left corner, and are clearly separated from the
bulk of the sample. Stars located below the dashed line in
Fig.~\ref{Fig:Sbsigma} may be supposed to be single (although some very
long-period binaries, with $P \ga 10$~y, may still hide among those). 
Their radial velocity dispersion suffers from a jitter which clearly
increases with increasing spectral line-width, as represented by the
parameter
$Sb$. For very large values of $Sb$ (in excess of about 9~\kms), the
diagram is populated almost exclusively by supergiants (star symbols in
the lower panel of Fig.~\ref{Fig:Sbsigma}). Many semi-regular and
irregular variables are located in the intermediate region, with $5 \la
Sb \la 9$~\kms. Strangely enough, spectroscopic
binaries seem to be lacking in this region. A more detailed discussion of
the properties of the binaries found among M giants is deferred to a
forthcoming paper (see Jorissen et al., 2004 for a preliminary account).

For K giants, no such structure is apparent in the  ($Sb,
\sigma_0(v_{r0}))$ diagram. It has been checked that the distribution of
stars along a $(N-1) [\sigma(v_{r0})/\epsilon]^2$ axis (where $N$ is the
number of measurements for a given star) follows a $\chi^2$
distribution, as expected (e.g., Jorissen \& Mayor 1988). This holds true
irrespective of the $Sb$ value, thus confirming the absence of structure
in the ($Sb, \sigma_0(v_{r0}))$ diagram. The binaries among K giants may thus
be identified by a straight $\chi^2$ test. A $1$\% threshold for the
first kind risk (of rejecting the null hypothesis that the star is not  a binary while actually true, {\it i.e.}, of considering the star as binary while actually single) has been chosen in the present study.

Among M giants, 42 binaries are found  (corresponding to an {\it observed}
frequency of spectroscopic binaries of 42/739, or 5.7\%). Among the 5952 K
giants, 817 spectroscopic binaries are found, corresponding to a frequency
of 13.7\%.  The large difference between these two frequencies
hints at some underlying physical cause, to be discussed in a forthcoming
paper (see Jorissen et al. 2004 for a preliminary
discussion). Most of the spectroscopic binaries (SBs) detected among our
samples of K and M giants are first detections. This large list of new
SBs constitutes an important by-product of the present
work. The new binaries are identified in Table A.1 (flags 0, 1, 5,
6 and 9  in column~24 of Table A.1) and will be the topic of a separate
paper. 

Among this total sample of 859 spectroscopic binaries, the
center-of-mass velocity could be computed (whenever the available
measurements were numerous enough to derive an orbit), 
estimated (in the case of low-amplitude orbits) or taken from the
literature 
for 216 systems
only, thus leaving 643 binaries which had to be discarded from the
kinematic study, because no reliable center-of-mass velocity could be
estimated.

\begin{figure}
   \centering
   \includegraphics[width=8cm]{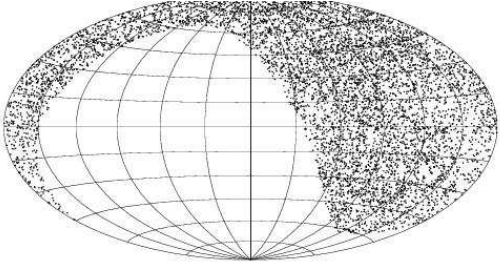}
      \caption{\label{Fig:Aitoff}
Distribution of the final sample on the sky, in galactic coordinates,
the galactic center being at the center of the map. The selection criterion
$\delta>0^\circ$ is clearly apparent on this map.
              }
   \end{figure}

After excluding those large-amplitude binaries as well as the dubious cases, 
5311 K giants and 719 M giants remain in the final
sample. Fig.~\ref{Fig:Aitoff} shows the distribution of our final sample on
the sky. Among this final sample of 6030 stars, 5397 belong
to the Hipparcos 'survey'.

\section{Kinematic analysis}
\label{Sect:kinematics}

In this section, we present the kinematic analysis of the data. Although
the velocity distribution is a function of position in the Galaxy, it will
be assumed here that our {\it local} stellar sample may be used to derive
the properties of the velocity distribution at the location of the Sun if
we correct the data from the effects of differential galactic rotation. 
First, we analyze the kinematics of the sample restricted to the 2774 stars with parallaxes accurate to
better than 20\%. In a second approach, designed to make full use of the 6030
available stars but without
being affected by the biases appearing when dealing with low-precision
parallaxes, the kinematic parameters are evaluated with a Monte-Carlo
method. Although easy to implement, the Monte Carlo method faces some
limitations (the parallaxes were drawn from a Gaussian distribution
centered on the {\it observed} parallax, not the true one as it
should). Therefore, in a third approach, we make use of a bayesian method
(the LM method; Luri et al. 1996), which allows us to derive simultaneously
maximum likelihood estimators of luminosity and kinematic parameters, and
which can identify possible groups present in the sample  by performing a
cluster analysis.     

In all cases, we correct for differential
Galactic rotation in order to deal with the true velocity of a star
relative to the Local Standard of Rest (the LSR is a reference frame in
circular orbit in the Galactic plane at the galactocentric distance of the
Sun). Given the observed values of the radial velocity $v_{r0}$ and the
proper motions $\mu_{l0}$ and $\mu_{b0}$, the corrected values are\
(Trumpler \& Weaver 1953):  

\begin{equation}
v_r = v_{r0} - A \, d \, {\rm cos}^2b \, {\rm sin}2l
\end{equation}
\begin{equation}
\kappa \, \mu_l = \kappa \,\mu_{l0} - A \, {\rm cos}2l - B
\end{equation}
\begin{equation}
\kappa \, \mu_b = \kappa \,\mu_{b0} + 1/2 \, A \, {\rm sin}2b \, {\rm sin}2l,
\end{equation}
\noindent where $\kappa$ is the factor to convert proper motions into
space velocities, and $d$ is the distance to the Sun.   
\noindent We used the values of the Oort constants $A$ and $B$ derived by
  Feast \& Whitelock\ (1997) from Hipparcos Cepheids, i.e. $A = 14.82
  \,{\rm km}\,{\rm s}^{-1}{\rm kpc}^{-1}$ and $B = -12.37 \,{\rm km}\,{\rm
  s}^{-1}{\rm kpc}^{-1}$.  The observed velocity is, after that correction,
  a combination of the peculiar velocity of the star and that of the Sun
  relative to the LSR.

\subsection{Analysis of the sample restricted to stars with the most
  precise parallaxes} 
\label{Sect:mostprecise}

The components of the
velocity of a star with respect to the Sun in the cartesian
coordinate system associated with galactic coordinates are the
velocity towards the galactic center $U$, the velocity in the
direction of Galactic rotation $V$, and the vertical velocity $W$. The
basic technique to calculate $U$, $V$ and $W$ is to invert the
parallax to estimate the distance and then use the proper motions and radial velocities as given by
Eqs. (1), (2) and (3). This simple procedure faces two major
difficulties: (i) the inverse parallax is a biased estimator of the distance  
(especially when the relative error on the parallax is high),
and (ii) the individual errors on the velocities 
cannot be derived from a simple first-order linear propagation of the
individual errors on the parallaxes. It is thus extremely
hard to estimate an error on the average velocities. If we limit
the sample to the stars with a relative parallax error smaller than 10$\%$,
we are left with 786 stars, which is too small a sample to analyze the general
behaviour of the K and M giants in the solar neighbourhood. We choose
instead to
restrict the sample to the stars with a relative parallax error smaller
than 20$\%$, in such a way that 2774 stars (2524 K and 250 M) remain. In
that case the classical first order approximation for the calculation of
the errors may still be applied (and the bias is very small, see
Brown et al. 1997), which yields from the errors on the proper motions, radial
velocities and parallaxes

\begin{equation}
\label{Eq:randomerror}
                        \begin{array}{l}
                            \langle \epsilon_{U} \rangle = 4.03 \,{\rm km}\,{\rm s}^{-1} \\
                           \langle \epsilon_{V} \rangle = 3.22 \,{\rm km}\,{\rm s}^{-1} \\
                            \langle \epsilon_{W} \rangle = 2.54 \,{\rm km}\,{\rm s}^{-1}.\\
                        \end{array}
\end{equation}

The contribution of the measurement errors to the uncertainty on the
sample mean velocity is $N^{-1/2}$ times the values given by~(\ref{Eq:randomerror}), where $N (=2774)$
is the sample size. This contribution is in fact negligible with respect to the ``Poisson noise'' (obtained
from the intrinsic velocity dispersion of the sample, see Eq.~(\ref{Eq:sigma20})). For the stars
with a relative parallax error smaller than 20$\%$, we obtain

\begin{equation}
\label{Eq:UVW20}
                        \begin{array}{l}
                            \langle U \rangle = -10.24 \pm 0.66 \,{\rm
                           km}\,{\rm s}^{-1} \\
                           \langle V \rangle = -20.51 \pm 0.43 \,{\rm km}\,{\rm s}^{-1} \\
                            \langle W \rangle = -7.77 \pm 0.34 \,{\rm
                           km}\,{\rm s}^{-1}. \\
                        \end{array}
\end{equation}

If there is no net radial and vertical motion at the solar position in the
Galaxy, we have hence estimated $U_\odot=-\langle U \rangle$ and
$W_\odot=-\langle W \rangle$ (see however Sect. 3.3.6).

Knowing that $\langle V \rangle$  is affected
by the asymmetric drift, which implies that the larger a stellar sample's velocity dispersion is,
the more it lags behind the circular Galactic rotation, it is interesting
to compare the mean value of $V$ for the K and M giants separately. For the
250 M giants, we obtain
\begin{equation}
\langle V \rangle = -23.42 \pm 1.48 \,{\rm km}\,{\rm s}^{-1}
\end{equation}
while for the 2524 K giants we obtain
\begin{equation}
\langle V \rangle = -20.22 \pm 0.44 \,{\rm km}\,{\rm s}^{-1}.
\end{equation}
This difference (at the $2\sigma$ level) between the two subsamples can be understood in terms of
the age-velocity dispersion relation. Indeed, the M giants must be a little
older than the K giants on average because only the low-mass stars can
reach the spectral type M on the Red Giant Branch and because the lifetime
on the main sequence is longer for lower mass stars. This implies that the subsample of M giants has a larger velocity
dispersion and rotates more slowly about the Galactic
Center than the subsample of K giants, in agreement with the asymmetric drift
relation. 

The velocity dispersion tensor is defined as $\langle {\bf (v - \langle v
  \rangle) \otimes (v - \langle v \rangle)} \rangle$. The diagonal components
  are the square of the velocity dispersions while the mixed components
  correspond to the covariances. For the stars 
with a relative parallax error smaller than 20$\%$, we obtain the classical
ordering for the diagonal components, i.e. $\sigma_U > 
\sigma_V > \sigma_W$:

\begin{equation}
\label{Eq:sigma20}
                        \begin{array}{l}
                         \sigma_U = 34.46 \pm 0.46 \,{\rm km}\,{\rm s}^{-1} \\
                         \sigma_V = 22.54 \pm 0.30 \,{\rm km}\,{\rm s}^{-1} \\
                         \sigma_W = 17.96 \pm 0.24 \,{\rm km}\,{\rm s}^{-1}. \\
                        \end{array}
\end{equation}
The asymmetric drift relation predicts a linear dependence of $\langle v \rangle = \langle V \rangle + V_\odot$ with the
radial velocity dispersion $\sigma_{U}^2$. If we adopt the peculiar
velocity of the sun from Dehnen \& Binney\ (1998), $V_\odot = 5.25
\,{\rm km}\,{\rm s}^{-1}$, we find for the full sample (K and M together)
\begin{equation}
\label{Eq:v}
\langle v \rangle =  -15.26 \,{\rm km}\,{\rm s}^{-1}.  
\end{equation}
On the other hand, if we adopt the parameter $k = 80 \pm 5 \,{\rm km}\,{\rm
  s}^{-1}$ from the asymmetric drift equation of Dehnen \& Binney\ (1998),
  we find 
\begin{equation}
\label{Eq:adr}
\langle v \rangle = - \sigma_{U}^2 / k = -14.9 \pm 1.3 \,{\rm km}\,{\rm
  s}^{-1}. 
\end{equation}
This independent estimate of $\langle v \rangle$ is in accordance with
(\ref{Eq:v}) and our values of $\langle V \rangle$ and $\sigma_{U}^2$ are
thus in good agreement with this value of $k$.

As we already noticed, the M giants are a little bit
older than the K giants on average and we thus expect the velocity
dispersions of the subsample of M giants to be higher. We find

\begin{equation}
\sigma_U({\rm M \, giants}) = 35.95 \pm 1.6 \,{\rm km}\,{\rm s}^{-1},
\end{equation}
and
\begin{equation}
\sigma_U({\rm K \, giants}) = 34.32 \pm 0.48 \,{\rm km}\,{\rm s}^{-1}.
\end{equation}
As expected, the difference is not very large, and the
radial velocity dispersions for the two subsamples satisfy the asymmetric drift
equation (\ref{Eq:adr}).

The mixed components of the velocity dispersion tensor involving vertical
motions vanish within their errors. Nevertheless, the mixed component in
the plane, that we denote $\sigma^2_{UV}$, is non-zero: this is not allowed
in an axisymmetric Galaxy and is further discussed in Sect. 3.3. We obtain

\begin{equation}
\sigma^2_{UV} = 134.26 \pm 13.28 \,{\rm km}^2\,{\rm s}^{-2}
\end{equation}
where the error corresponds to the 15 and 85 percentiles of the correlation
coefficient, assuming that the sample of $U,V$ velocities is drawn from a
two-dimensional gaussian distribution. 

\begin{figure}
   \centering
   \includegraphics[width=10cm]{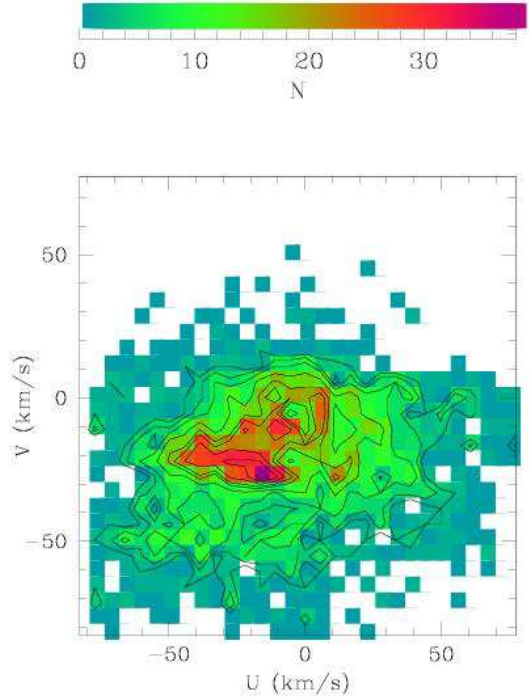}
      \caption{\label{Fig:UV}
Density of stars with precise
parallaxes ($\sigma_\pi/\pi \le 20$\%) in the $UV$-plane. The colours indicate the number of stars
in each bin. The contours indicate the bins with  3, 4, 7, 12, 17, 20, 25,
30, 35 and 40 stars respectively. The concentration of stars
around $U \simeq -35$~\kms, $V \simeq -45$~\kms\ contributes largely to the vertex
deviation, while other peaks already identified by Dehnen\ (1998) at $U=-40$~\kms,
$V=-25$~\kms\ (Hyades supercluster), at $U=10$~\kms, $V=-5$~\kms\ (Sirius
moving group), and at $U=-15$~\kms,
$V=-25$~\kms\ (Pleiades supercluster) are also present.
              }
   \end{figure}

In order to parametrize the deviation from dynamical axisymmetry, a useful quantity
is the vertex deviation $l_v$, i.e., the angle one has to rotate the
$(U,V)$ coordinate system in order to diagonalize the velocity dispersion
tensor. The value of the vertex deviation for our sample is 
\begin{equation}
l_v \equiv 1/2 \, {\rm arctan}\,(2\sigma^2_{UV}/(\sigma^2_U - \sigma^2_V))
= 10.85^\circ \pm 1.62^\circ.
\end{equation}
This vertex deviation for giant stars is not in accordance with a
perfectly axisymmetric Galaxy and could be caused by non-axisymmetric
perturbations in the solar neighbourhood, or by a deviation from
equilibrium (i.e. moving groups due to inhomogeneous star formation,
following the classical theory of moving groups). A hint to the
true nature of this vertex deviation could be the local anomaly in the
$UV$-plane, the so-called $u$-anomaly\ (e.g. Raboud et al. 1998). If we
calculate the mean velocity $\langle U \rangle$ of the 
stars with $V<-35 \,{\rm km}\,{\rm s}^{-1}$, we find that it is
largely negative ($\langle U \rangle = -22\,{\rm km}\,{\rm s}^{-1}$). It
denotes a global outward radial motion of the stars that lag behind the galactic rotation. We see on Fig.~\ref{Fig:UV} that this anomaly is due to a
clump located at $U \simeq -35$~\kms, $V \simeq -45$~\kms, the already
known 'Hercules' stream\ (Fux
2001). On the other hand, Fig.~\ref{Fig:UV} reveals in fact a rich
small-scale structure in the $UV$-plane, with several clumps which can be associated with
known kinematic features: we clearly see small peaks at $U=-40$~\kms, $V=-25$~\kms\ (corresponding to the Hyades
supercluster), at $U=10$~\kms, $V=-5$~\kms\ (Sirius moving group), and at $U=-15$~\kms,
$V=-25$~\kms\ (Pleiades supercluster). More precise values for these peaks
will be given in Sect. 3.3 where their origin will also be discussed.

\subsection{Monte Carlo simulation}
\label{Sect:MonteCarlo}

To reduce the bias introduced by the non-linearity of the parallax-distance
transformation, the results discussed in the previous Section were obtained
by restricting the sample to stars with a relative parallax error smaller
than 20$\%$. In this Section, we use all the available stars, without any
parallax truncation, but perform a Monte-Carlo simulation to properly
evaluate the errors and biases on the kinematic parameters. 

We constructed synthetic samples by drawing the stellar parallax
from a gaussian distribution centered on the {\it observed} parallax value
for the corresponding star, and with a dispersion corresponding to the
uncertainty on the observed parallax. Note that this procedure is not
strictly correct, as the gaussian distribution should in fact be centered
on the (unknown) {\it true} parallax. To overcome this difficulty, a fully
bayesian approach will be applied in yet another analysis of the data, as
described next in Sect.~\ref{Sect:LM}.    

Some {\it a priori} information has nevertheless been included to truncate
the gaussian parallax distribution in the present Monte Carlo approach,
since the giant stars in our sample should not be brighter than $M_{Hp} =
-2.5$ (see Figs.~\ref{Fig:dhrHyPl}, \ref{Fig:dhrSi}, \ref{Fig:dhrHe},
\ref{Fig:dhrB}). That threshold has been increased to $M_{Hp} = -5$ (see
Fig.\ref{Fig:dhrY}) for the stars
flagged as supergiants. This prescription thus corresponds to assigning a
minimum admissible parallax $\pi_{\rm min}$ to any given star. It prevents
that very small parallaxes drawn from the gaussian distribution yield
unrealistically large space velocities. The parallax distribution used in
the Monte Carlo simulation thus writes

\begin{eqnarray}
                          P(\pi) =                
\frac{1}{(2\Pi)^{1/2}\sigma_{\pi_{obs}}}
\exp\left(-\frac{1}{2}\left(\frac{\pi-\pi_{obs}}{\sigma_{\pi_{obs}}}\right)^2\right)
                          & \rm{if} & \pi \geq \pi_{\rm min} \\
                          P(\pi) = 0
                          & \rm{if} & \pi < \pi_{\rm min}
                          \end{eqnarray}

The mean space velocities and the velocity dispersions are calculated for
each simulated sample, and finally we adopt the average of these values
(of the mean velocities and of the velocity dispersions) over 4000
simulated samples as the best estimate of the true kinematic parameters. 
We thus obtain: 

\begin{equation}
\label{Eq:UVWMC}
                        \begin{array}{l}
                            \langle U \rangle = -10.25 \pm 0.15 \,{\rm
                           km}\,{\rm s}^{-1} \\
                           \langle V \rangle = -22.81 \pm 0.15 \,{\rm km}\,{\rm s}^{-1} \\
                            \langle W \rangle = -7.98 \pm 0.09 \,{\rm
                           km}\,{\rm s}^{-1}. \\
                        \end{array}
\end{equation}

If there is no net radial and vertical motion in the solar neighbourhood
(see however Sect. 3.3.6), we may write $U_\odot=-\langle U \rangle$ and 
$W_\odot=-\langle W \rangle$. The results given by (\ref{Eq:UVWMC}) are very close to
those estimated from stars with relative errors on the parallax smaller than 20\%  (see
(\ref{Eq:UVW20})). They are in agreement with the
values derived by Dehnen \& Binney\ (1998) on the basis of Hipparcos proper
motions of main
sequence stars ($U_{\odot} = 10.00 \pm 0.36 \,{\rm km}\,{\rm s}^{-1}$,
$W_{\odot} = 7.17 \pm 0.38 \,{\rm km}\,{\rm s}^{-1}$). The value of
$U_\odot$ is not in perfect accordance with the one
derived by Brosche et al.\ (2001), who found $U_{\odot} = 9.0 \pm 0.5
\,{\rm km}\,{\rm s}^{-1}$ from photometric distances and Hipparcos proper
motions of K0-5 giants, nor with the one derived by Zhu\ (2000) who found
$U_{\odot} = 9.6 \pm 0.3 \,{\rm km}\,{\rm s}^{-1}$ with the same stars as
ours but without the radial velocity data. The value of $W_\odot$
contradicts slightly the one
derived by Bienaym\'e\ (1999), who found $W_{\odot} = 6.7 \pm 0.2 \,{\rm
  km}\,{\rm s}^{-1}$ from Hipparcos proper motions. Nevertheless, these
considerations on the solar motion are not very useful since we stress
in Sect. 3.3.6 that the motion of the Sun is
difficult to derive anyway because there are conceptual uncertainties on the
mean motion of stars in the solar neighbourhood.  

For the velocity dispersions, we find

\begin{equation}
                        \begin{array}{l}
                         \sigma_U = 40.72 \pm 0.58 \,{\rm km}\,{\rm s}^{-1} \\
                         \sigma_V = 32.23 \pm 1.41 \,{\rm km}\,{\rm s}^{-1} \\
                         \sigma_W = 22.55 \pm 0.95 \,{\rm km}\,{\rm s}^{-1}. \\
                        \end{array}
\end{equation}

These velocity dispersions are somewhat larger than those found for the
restricted sample (\ref{Eq:sigma20}), which is not surprising since 60\% of the high-velocity stars (as identified in Sect.~3.3.6) are not present in the sample restricted to the most precise parallaxes. 
%It could also
%be a slight effect of the larger volume of the Galaxy probed
%by the complete sample.  
Regarding the asymmetric drift and the slightly
larger age of the M giants, the Monte Carlo method and the analysis of
the restricted sample reach the same conclusion.

Concerning the mixed component of the velocity dispersion tensor in the
plane, we find
\begin{equation}
\sigma^2_{UV} = 188.23 \pm 40.9 \,{\rm km}^2\,{\rm s}^{-2}
\end{equation}
which leads to a clearly non-zero vertex deviation of
\begin{equation}
l_v \equiv 1/2 \, {\rm arctan}\,(2\sigma^2_{UV}/(\sigma^2_U - \sigma^2_V))
= 16.2^\circ \pm 5.6^\circ.
\end{equation}

Interestingly, Soubiran et al. (2003) found no vertex deviation for
low-metallicity stars in the disk, 
and concluded that this was consistent with an axisymmetric Galaxy: this is
absolutely not the case for our sample of late-type giants. Furthermore,
the value we find for the vertex deviation is larger than the one
derived for late-type stars by Dehnen \& Binney\ (1998), who showed that
the vertex deviation drops from $30^\circ$ for young stellar populations
(probably due to young groups concentrated near the origin of the
$UV$-plane, see also Sect.~3.3.7) to a constant value of $10^\circ$ for older populations. Bienaym\'e\ (1999)
also found from Hipparcos proper motions a vertex deviation of $9.2^\circ$
for the giant stars. We conclude from our sample that the vertex deviation
is significantly non-zero: we suggest a possible origin for this vertex
deviation (which could be the same origin as for the vertex deviation of
younger populations) in Sect. 3.3.7. 

\subsection{Bayesian approach}
\label{Sect:LM}

To obtain the kinematic characteristics of our sample in a more
rigorous way, we have
applied the Luri-Mennessier (LM) method described in detail by Luri (1995)
and Luri et al. (1996).  
The starting point of this method is a model describing the basic
morphological characteristics we can safely expect from the
sample (spatial distribution, kinematics and absolute magnitudes), and a
model of the selection criteria used to define the
sample. These models are used to build a distribution function (DF)
intended to describe the observational characteristics of the sample. The {\it
a priori} distribution function adopted is a linear combination of
partial DFs, each of which describes a group of stars (called ``base
group''). Each partial DF combines a kinematic model
(the velocity ellipsoid introduced by Schwarzschild 1907) with a gaussian
magnitude distribution, and an exponential height distribution uncorrelated
with the velocities. The phenomenological 
model adopted is obviously not completely rigorous, but has
the advantage of being able to identify and quantify the different subgroups
present in the data and possibly related to extremely complex dynamical
phenomema, which cannot be easily parametrized.  

The values of the parameters of the DF can be determined
from the sample using a bayesian approach: the model is adjusted to the
sample by a maximum likelihood fit of the parameters. The values of the
parameters so obtained provide the best representation of the sample given the
{\it a priori} models assumed.

In the following Sects.~3.3.1 and 3.3.2 we describe the ingredients
of the models used in this paper. Thereafter, in Sects.~3.3.3 to 3.3.6, we present the results of the maximum likelihood fit, and we discuss
in Sect.~3.3.7 the different possible physical interpretations of
those results. 

%The method we have decided to use in this section is completely bayesian
%(the LM method of Luri et al. 1996).  It assumes that the considered sample
%is a mixture of several base groups. Each base group is described by a
%distribution function (DF): 

%\begin{equation}
%                          \begin{array}{l}
%                          f(M,d,l,b,U,V,W) =\\
% {\rm exp}\left( -\frac{1}{2}(\frac{U-\langle
% U\rangle}{\sigma_U})^2-\frac{1}{2}(\frac{V-\langle
% V\rangle}{\sigma_V})^2-\frac{1}{2}(\frac{W-\langle
% W\rangle}{\sigma_W})^2-\frac{1}{2}(\frac{M- \langle M \rangle}{\sigma_M})^2\\
%-\frac{d \; \sin b}{z_0} \right) d^2 \; \cos b,
%                          \end{array}
%\end{equation}
%where $b$ denotes the galactic latitude, $z_0$ the scale height of the
%group, $M$ the absolute magnitude, and $d$ the distance from the Sun. A
%vertex deviation is also introduced. 

%To find the parameters of the DF
%for each base group, we use a maximum likelihood method, incorporating the
%selection criteria of the sample, the individual errors of the observed values
%for each star and the interstellar absorption (absorption model of Arenou
%et al. 1992). 

%----------------------------------------------
% Distribution functions of the physical model
%----------------------------------------------
\subsubsection{Phenomenological model}

To build a phenomenological model of our sample, we assume that it is a mixture of 
stars coming from several ``base groups''. A given group represents a fraction
$w_i$ of the total sample and its characteristics are described by
the following components: 

\begin{itemize}
\item {\it Spatial distribution:} An exponential disk of scale height $Z_0$ in the
direction perpendicular to the galactic plane

       \begin{equation}
          \varphi_{\rm e}(d,l,b) = \exp \left(
                                      - \frac{|d \sin b|}{Z_0}
                                     \right)
          \; d^2 \cos b
       \end{equation}
       
       and a uniform distribution along the galactic plane (a realistic 
       approximation for samples in the solar neighbourhood like ours).

\item {\it Velocity distribution:} A Schwarzschild ellipsoid for the velocities
of the stars with respect to their Local Standard of Rest (LSR):

       \begin{equation}
       \begin{array}{l}
         \varphi_{\rm v}(U',V',W) = e^ {
                                  - \frac{1}{2}
                                  \left(
                                         \frac{U'}{\sU'}
                                  \right)^2
                                  - \frac{1}{2}
                                  \left(
                                         \frac{V'}{\sV'}
                                  \right)^2 
                                  - \frac{1}{2}
                                  \left(
                                         \frac{W-W_0}{\sW}
                                  \right)^2
                                }
       \end{array}
       \end{equation}

where
\begin{equation}
                        \begin{array}{l}
                        U'= (U-U_0) \cos l_v - (V-V_0) \sin l_v \\
                        V'= (U-U_0) \sin l_v + (V-V_0) \cos l_v
                        \end{array}
\end{equation}
and where $l_v$ is the vertex deviation.

\item {\it Galactic rotation:} An Oort-Lindblad rotation model at first order, 
where the rotation velocity is added to the mean LSR velocity (given by the
Schwarzschild ellipsoid above). The same values were adopted for the Oort
constants as in Sects. 3.1 and 3.2

\item {\it Luminosity:} We have adopted a gaussian distribution for the
  absolute magnitudes of the stars:  

      \begin{equation}
         \varphi_{\rm M}(M) = \exp \left( - \frac{1}{2}
                                    \left(
                                        \frac{M-M_0}{\sM}
                                    \right)^2
                                    \right).
      \end{equation}

This is just a first rough approximation. A better model would include
a dependence of the absolute magnitude on a colour index but is more
complex (we leave it for future papers).  

\item {\it Interstellar absorption:} In the LM method, the correction of
  interstellar absorption is integrated in the formalism. A model, giving
  its value as a function of the position  $(d,l,b)$, is needed: the Arenou
  et al. (1992) model has been chosen here. 

\end{itemize}

The DF of each base group is simply the product of the
space, velocity and magnitude distributions presented here, with different
values for the model parameters. The total distribution function $\D$ of
the sample is a linear combination of the partial DFs of the
different base groups.

Both the relative fractions of the different base groups and
the model parameters will be determined using a Maximum Likelihood fit, as
described in Sect. ~\ref{Sect:ML}. However, the number of groups ($n_g$)
composing our sample is not known {\it a priori}. A likelihood test -- like
Wilk's test, Soubiran et al.\ (1990) -- will be used to determine it: ML
estimations are performed with $n_g=1,2,3,\ldots$  and the maximum  likelihoods
obtained for each case are compared using the test to decide on the most likely
value of $n_g$.  

\subsubsection{Observational selection and errors}

As pointed out above, the correct description of the observed
characteristics of the sample requires that its selection criteria be included in
the model. In our case, our sample of giants (like the 
Hipparcos Catalogue as a whole) is composed primarily of stars belonging 
to the Hipparcos {\it survey} plus stars added on the basis of several
heterogeneous criteria. 

In the case of the Hipparcos survey, the selection criteria are just based on the
apparent $V$ magnitude.  For a given line of sight, 
with galactic latitude $b$, the survey is complete up to 

\begin{equation}
  V = 7.3 + 1.1 |\sin b|.
\end{equation} 

In our case, the $H_p$ Hipparcos magnitude has been used instead of
$V$. Therefore the survey stars in our sample will follow a similar
``completeness law'' in $H_p$ that we have adopted to be approximately 

\begin{equation}
  H_p = 7.5 + 1.1 |\sin b|
\end{equation} 
assuming an average $H_p - V$ of $0.2$ mag for the stars in our sample.

This law, altogether with the cutoff in declination ($\delta > 0^\circ$),
allows us to quite realistically model the selection of the
survey stars in the sample, but  leaves out 10\% 
of the total sample. In order to be able to use the full sample, we must
define the selection of the non-survey stars as well, as (approximately)
done by the following completeness law:

\begin{itemize}
  \item Completeness linearly decreasing from 1 at $H_p = 7.5 + 1.1 |\sin b|$
  down to zero at  $H_p=11.5$ 
\end{itemize} 
This condition together with the completeness up to $H_p = 7.5 + 1.1 |\sin
b|$ , and with the cutoff in declination, defines the
selection of the complete sample. 

The individual errors on the astrometric and photometric data are also
taken into account in the model. We assume that the observed values are
produced by gaussian distributions of observational errors around the
``true'' values, with standard deviations given by the errors quoted in the
Hipparcos Catalogue, the Tycho-2 catalogue or the CORAVEL database. 

In the end, we can define for each star a joint distribution function of the
true and observed values:     
\begin{equation}
\mu(\x,\z \| \t) = \D(\x \| \t) \, \epsilon (\z \| \x) \, S(\z)
\end{equation}
where $\x$ are the true values (position in the LSR, velocities with
respect to the LSR, absolute magnitude), $\z$ the observed values
(position on the sky, parallax, proper motions, radial velocity, apparent
magnitude), $\t$ the set of parameters of the model, $S$ the selection
function taking into account the selection criteria of the sample, and $\epsilon$
the gaussian distribution of observational errors.

For each star, we can then easily deduce the distribution function of the
observed values:  
\begin{equation}
\O(\z \| \t) = \int \mu(\x,\z \| \t) \: \d\x
\end{equation}

The fact that the selection function and the error
distribution are taken into account in this distribution function
prevents from all the possible biases. 

\subsubsection{Maximum likelihood}
\label{Sect:ML}

The principle of maximum likelihood (ML) can be briefly described as
follows: let $\vec{z}$  be the random variable of the observed values
following the density law given by $\O (\z \| \t_0)$, where $\t_0 = ( w_1,
M_{0_1}, \sigma_{M_1}, U_{0_1}, \sigma_{U'_1}, V_{0_1}, \sigma_{V'_1},
W_{0_1}, \sigma_{W_1}, Z_{0_1}, l_{v_1}, w_2, \ldots, l_{v_{n_g}})$ is the
set of unknown parameters on which it depends.   

The {\em Likelihood Function} is defined as

  \begin{equation}  
     \L (\t) = \prod_{i=1}^{n_*} \O (\z_i \| \t) .
  \end{equation}

The value of $\t$ which maximizes this function is the ML estimator, $\t_\ML$,
of the parameters $\t_0$ characterizing the density law of the sample.
It can be shown that $\t_\ML$ is asymptotically non-biased, asymptotically
gaussian and that for large samples, it is the most efficient estimator (see
Kendall \& Stuart 1979).

Once the ML estimator of the parameters has been found, simulated samples
(generation of random numbers following the given distribution) are used to
check the equations and programs developed for the estimation. They also allow
a good estimation of the errors on the results and the 
detection of possible biases: once a \ML ~estimation $\t_\ML$ has been
obtained,  several samples corresponding to the parameters $\t_\ML$ are
simulated and the method
is applied to them. The comparison of the results (which could be
called ''the estimation of the estimation'') with the original $\t_\ML$
allows us to detect possible biases. The dispersion of these
results $\sigma(\t_{\ML})$ can be taken as the error of the estimation.

The low maximum likelihood obtained when the full sample is modelled with a
single base group indicates that the kinematic properties of giant stars in the solar neighbourhood 
cannot be fitted by a single Schwarzschild ellipsoid. The first acceptable
solution requires three base groups: one of bright giants and supergiants
with 'young' kinematics (further discussed in Sect. ~\ref{Sect:groups}), 
one of high-velocity stars and finally one group of 'normal' stars. This
third group exhibits plenty of
small-scale structure, and this small-scale structure can be successfully
modelled by 3 more base groups (leading to a total of 6 groups in the
sample, called groups Y, HV, HyPl, Si, He, and B, further discussed in
Sect. 3.3.6).
These supplementary groups are statistically significant, as revealed by a
Wilks test. Statistically, solutions with 7 or 8 groups are even better,
but these solutions are not stable anymore and are thus not useful (they
depend too much on the observed values of some individual stars). Table 1
lists the values of the ML parameters for a model using just the 5177 (out
of 5397) survey stars which comply with the completeness law (Eq.~(26)). Table 2
lists those values for a model using the whole sample of 6030 stars with our modified
selection law. The models for the survey stars alone and for the whole
sample give very similar results (except for the scale height $Z_0$ of
group HV -- see Sect. 3.3.6 -- which is not well constrained since our
sample does not go far enough above the galactic plane), thus providing a
strong indication that the results for the whole sample are reliable. The
errors on the parameters listed in Tables 1 and 2 are the dispersions
coming from the simulated samples. 

\begin{table*}
\caption{Maximum-Likelihood parameters obtained using 5177 survey
  stars only. The number of stars belonging to each group (as listed in column~34 of Table~A.1)  is given in the last row. Although obtained from the assignment process described in  Sect.~\protect\ref{Sect:group}, it is {\it not,} strictly, a ML-parameter, in contrast to  the `fraction' $w_i$ ($i=1,...6)$ of the
   whole sample belonging to the considered group, as listed in the previous row. Therefore, 
  the observed fraction of stars in each group could be slightly different from the
  true fraction $w_i$ of that group in the entire population.
  %, due to
  %to some selection biases. Those biases have
  %been taken into account when determining the ML parameters from the joint
  %distribution of Eq.(27).
  }
\begin{tabular}{l| l | l | l | l | l | l }
\hline
  & group Y & group HV & group HyPl & group Si & group He & group B \\
\hline
$M_0$ (mag) & 0.61 $\pm$  0.91 & 2.21 $\pm$ 0.22 & 1.15 $\pm$
  0.37 & 1.08  $\pm$ 0.57 & 1.27 $\pm$ 0.37 &  1.23 $\pm$ 0.17 \\
$\sigma_M$ (mag) & 1.85 $\pm$ 0.24 & 1.29 $\pm$ 0.10  & 0.88 $\pm$ 0.21 &
  1.16 $\pm$ 0.24 & 0.88 $\pm$ 0.12 & 1.05 $\pm$ 0.08 \\
$U_0$ (\kms)& -11.63 $\pm$ 2.30 & -17.56 $\pm$ 3.46  & -31.23
  $\pm$ 0.99 & 5.19 $\pm$ 3.03 & -42.09 $\pm$ 4.76 & -2.92 $\pm$ 1.54 \\
$\sigma_{U'}$ (\kms) & 15.61 $\pm$ 1.22 &  53.45 $\pm$ 2.14 & 11.46 $\pm$
  1.53 & 13.48 $\pm$ 2.66 & 26.08 $\pm$ 13.13 & 31.82 $\pm$ 1.56 \\
$V_0$ (\kms) & -11.57 $\pm$ 2.08 & -43.96 $\pm$ 3.20 & -20.00 $\pm$
  0.75 & 4.22  $\pm$ 1.95 & -50.60  $\pm$ 4.67 & -15.15 $\pm$ 2.36\\
$\sigma_{V'}$ (\kms) & 9.44 $\pm$ 1.36 & 36.14 $\pm$ 1.79 & 4.89 $\pm$
  1.10 & 4.59 $\pm$ 4.22 & 8.62 $\pm$ 3.03 & 17.60 $\pm$ 0.78\\
$W_0$ (\kms) & -7.79 $\pm$ 0.95 & -7.76 $\pm$ 2.30 & -4.82 $\pm$
  1.85 & -5.57 $\pm$ 2.13 & -6.94 $\pm$ 3.56 &  -8.16 $\pm$ 0.91 \\
$\sigma_W$ (\kms) & 6.94 $\pm$ 0.78 & 32.54 $\pm$ 1.54 & 8.76 $\pm$
  1.08 & 9.43 $\pm$ 2.23 & 16.66 $\pm$ 1.62 & 16.26 $\pm$ 0.77 \\
$Z_0$ (pc) & 73.44 $\pm$ 6.33 & 901.04 $\pm$ 440.13 & 128.70 $\pm$
  19.21 & 149.43 $\pm$ 23.67 & 201.80 $\pm$ 55.24 & 196.05 $\pm$
  12.34 \\ 
$l_v$ ($^\circ$) & 17.07 $\pm$ 5.81 & 0.22 $\pm$ 4.89 & -5.65 $\pm$
  4.59 & -14.17 $\pm$ 25.28 & -5.67 $\pm$ 11.16 & -0.22 $\pm$ 5.21 \\
fraction $w$ & 0.086 $\pm$ 0.013 & 0.149 $\pm$ 0.022 & 0.071 $\pm$ 0.007 & 0.050
  $\pm$ 0.030 & 0.065 $\pm$ 0.017 & 0.579 $\pm$ 0.047\\
number of stars & 345 & 505 & 334 & 204 & 372 & 3417\\

\hline

\end{tabular}
\end{table*}

\begin{table*}
\caption[]{Same as Table 1 for the full sample of 6030 stars}
\begin{tabular}{l| l | l | l | l | l | l }
\hline
  & group Y & group HV & group HyPl & group Si & group He & group B \\
\hline
$M_0$ (mag) & 0.68 $\pm$  0.25 & 2.04 $\pm$ 0.23 & 0.95  $\pm$ 0.13 &
  0.87 $\pm$ 0.19 &  1.21 $\pm$ 0.18 & 1.04 $\pm$ 0.07 \\
$\sigma_M$ (mag) & 1.78 $\pm$ 0.06 & 1.43 $\pm$ 0.08  & 0.74 $\pm$ 0.10 &
  1.12 $\pm$ 0.08 & 0.95 $\pm$ 0.09 & 1.07 $\pm$ 0.03 \\
$U_0$ (\kms) & -10.41 $\pm$ 0.94 & -18.50 $\pm$ 2.82  & -30.34
  $\pm$ 1.54 & 6.53 $\pm$ 1.93 & -42.13 $\pm$ 1.95 & -2.78 $\pm$ 1.07 \\
$\sigma_{U'}$ (\kms) & 15.37 $\pm$ 1.09 &  58.02 $\pm$ 1.87 & 11.83 $\pm$
  1.34 & 14.37 $\pm$ 2.05 & 28.35 $\pm$ 1.68 & 33.30 $\pm$ 0.70 \\
$V_0$ (\kms) & -12.37 $\pm$ 0.90 & -53.30 $\pm$ 3.10 & -20.27 $\pm$
  0.62 & 3.96  $\pm$ 0.67 & -51.64  $\pm$ 1.07 & -15.42 $\pm$ 0.82\\
$\sigma_{V'}$ (\kms) & 9.93 $\pm$ 0.75 & 41.36 $\pm$ 1.70 & 5.08 $\pm$
  0.76 & 4.63 $\pm$ 0.71 & 9.31 $\pm$ 1.22 & 17.94 $\pm$ 0.80\\
$W_0$ (\kms) & -7.75 $\pm$ 0.57 & -6.61 $\pm$ 1.82 & -4.82 $\pm$
  0.80 & -5.80 $\pm$ 1.15 & -8.06 $\pm$ 1.30 &  -8.26 $\pm$ 0.38 \\
$\sigma_W$ (\kms) & 6.69 $\pm$ 0.58 & 39.14 $\pm$ 1.69 & 8.75 $\pm$
  0.74 & 9.66 $\pm$ 0.82 & 17.10 $\pm$ 1.63 & 17.65 $\pm$ 0.34 \\
$Z_0$ (pc) & 80.27 $\pm$ 6.18 & 208.08 $\pm$ 27.89 & 106.27 $\pm$ 14.37 &
  127.96 $\pm$ 19.65 & 132.89 $\pm$ 9.42 & 141.21 $\pm$ 5.20 \\
$l_v$ ($^\circ$) & 16.40 $\pm$ 10.26 & 0.16 $\pm$ 4.77 & -8.79 $\pm$
  4.05 & -11.96 $\pm$ 3.53 & -6.53 $\pm$ 2.77 & -2.18 $\pm$ 1.60 \\
fraction $w$ & 0.096 $\pm$ 0.008 & 0.106 $\pm$ 0.010 & 0.070 $\pm$ 0.008 & 0.053
  $\pm$ 0.009 & 0.079 $\pm$ 0.009 & 0.596 $\pm$ 0.015 \\  
number of stars & 413 & 401 & 392 & 268 & 529 & 4027\\

\hline

\end{tabular}
\end{table*}

%\begin{figure}
%\centering
%   \includegraphics[width=8cm]{Sample_Mua.eps}
%\caption[]{\label{Fig:simmualpha}
  %}
%\end{figure}

%\begin{figure}
%\centering
%   \includegraphics[width=8cm]{Sample_Vr.eps}
%\caption[]{\label{Fig:simvr}
  
%}
%\end{figure}

\subsubsection{Group assignment}
\label{Sect:group}

The probabilities for a given star to belong to the various groups is
provided by the LM method, and the most probable group along with the
corresponding probability is listed in Table A.1. Indeed, let $w_j$ be the
{\it a priori} probability (i.e. the ML parameter of Table 2) that a star belongs to the $j$th group and $\O_j (\z \; | \; \t_j )$ the distribution of the
observed quantities $\z$ in this group (deduced from the phenomenological model
adopted and depending on the parameters $\t_j$ of this model for the group). 
Then, using Bayes formula, the  {\it a posteriori} probability for a 
star to belong to the $j$th group given its measured values $\z_*$ is:

\begin{equation}
     P( * \in G_j \: | \: \z_*) =
     \frac{
           w_{j} \O( \z_* |  \t_{j}  )
          }
          {
            \sum_{k=1}^{n_g} w_{k}
            \O_k( \z_* |  \t_{k}  )
          } .
\end{equation}

\noindent Using this formula the {\it a posteriori} probabilities that the
star belongs to a given group can be compared, and the star can be assigned
to the most likely one (Table A.1).

Note that this procedure, like any method of statistical classification,
will have a certain percentage of misclassifications. However, the reliability
of each assignment is clearly indicated by the probability given in column 35 of Table~A.1.

\subsubsection{Individual distance estimates}

Once a star has been assigned to a group, and given the \ML ~estimator
of the group parameters and the observed
values for the star $\z_*$, one can obtain the marginal
probability density law $\R (d)$ for the distance of the star from the global
probability density function. \\

\noindent This can then be used to obtain the expected value of the distance

\begin{equation}
  \overline{d} = \int_0^{\infty} d \: \R (d) \: \d d
\end{equation}

\noindent and its dispersion

\begin{equation}
\label{Eq:errdist}
  \epsilon_d^2 = \int_0^{\infty} (d - \overline{d})^2 \: \R (d) \: \d d .
\end{equation}

\begin{figure}
\centering
   \includegraphics[width=9cm]{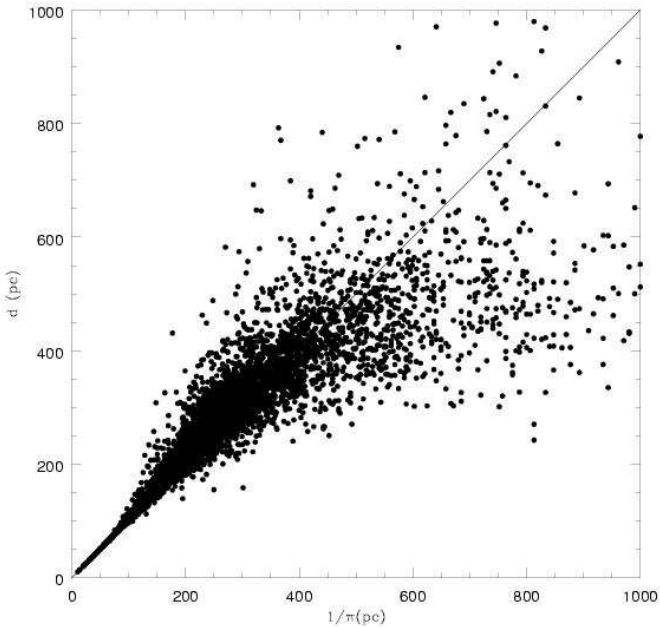}
\caption[]{\label{Fig:LMdistances}
Comparison of the (biased) distances obtained from a simple inversion
of the  parallax, and the maximum-likelihood distances $\overline{d}$
obtained from the LM method.  
}
\end{figure}

\noindent The first can be used as a distance estimator free from biases and the
second as its error (Columns 28 and 29 of Table A.1). Fig. \ref{Fig:LMdistances} reveals that the biases in
the distance derived from the inverse parallax are a combination of
those resulting from truncations in parallax and apparent magnitude as
discussed in Luri \& Arenou\ (1997).

\subsubsection{The various groups present in the sample}
\label{Sect:groups}

\begin{figure}
   \centering
   \includegraphics[width=8.5cm]{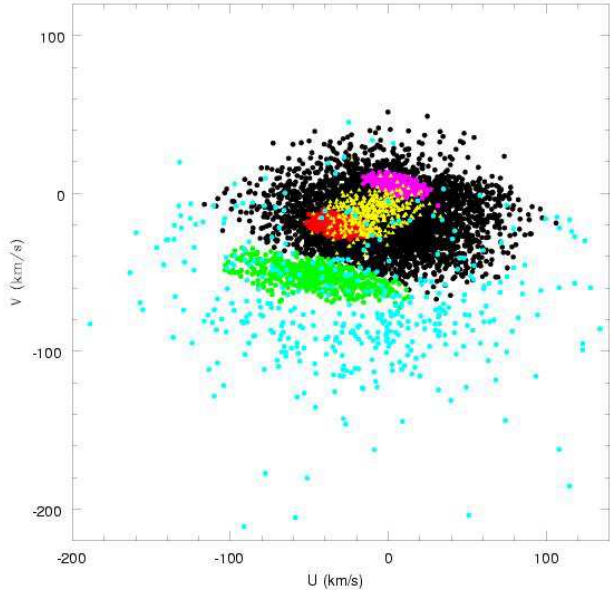}
      \caption{\label{Fig:UVluri}
      All the stars plotted in the $UV$-plane with their values of $U$ and
      $V$ deduced from the LM method. The 6 different groups are
      represented on this figure: group Y in yellow, group HV in blue,
      group HyPl in red, group Si in magenta, group He in green and group B
      in black. Note that the yellow group (Y) extends just far enough to
      touch both the red (HyPl) and magenta (Si) groups.}
   \end{figure}

We have plotted in Fig.~\ref{Fig:UVluri} the stars in the $UV$-plane by
using the expected values of $U$ and $V$ deduced from the LM method. The 6
different groups are represented on this figure. The structure of the
$UV$-plane is similar to the one already identified (in
Fig.~\ref{Fig:UV}) for the stars with precise parallaxes ($\sigma_\pi/\pi
\le 20$\%). This similarity is a strong indication that the subgroups
identified by the LM method are
not artifacts of this method. Several groups can be identified with
known kinematic features of the solar neighbourhood.

\noindent {\it Group Y: The young giants} 

\begin{figure}
   \centering
   \includegraphics[width=8.5cm]{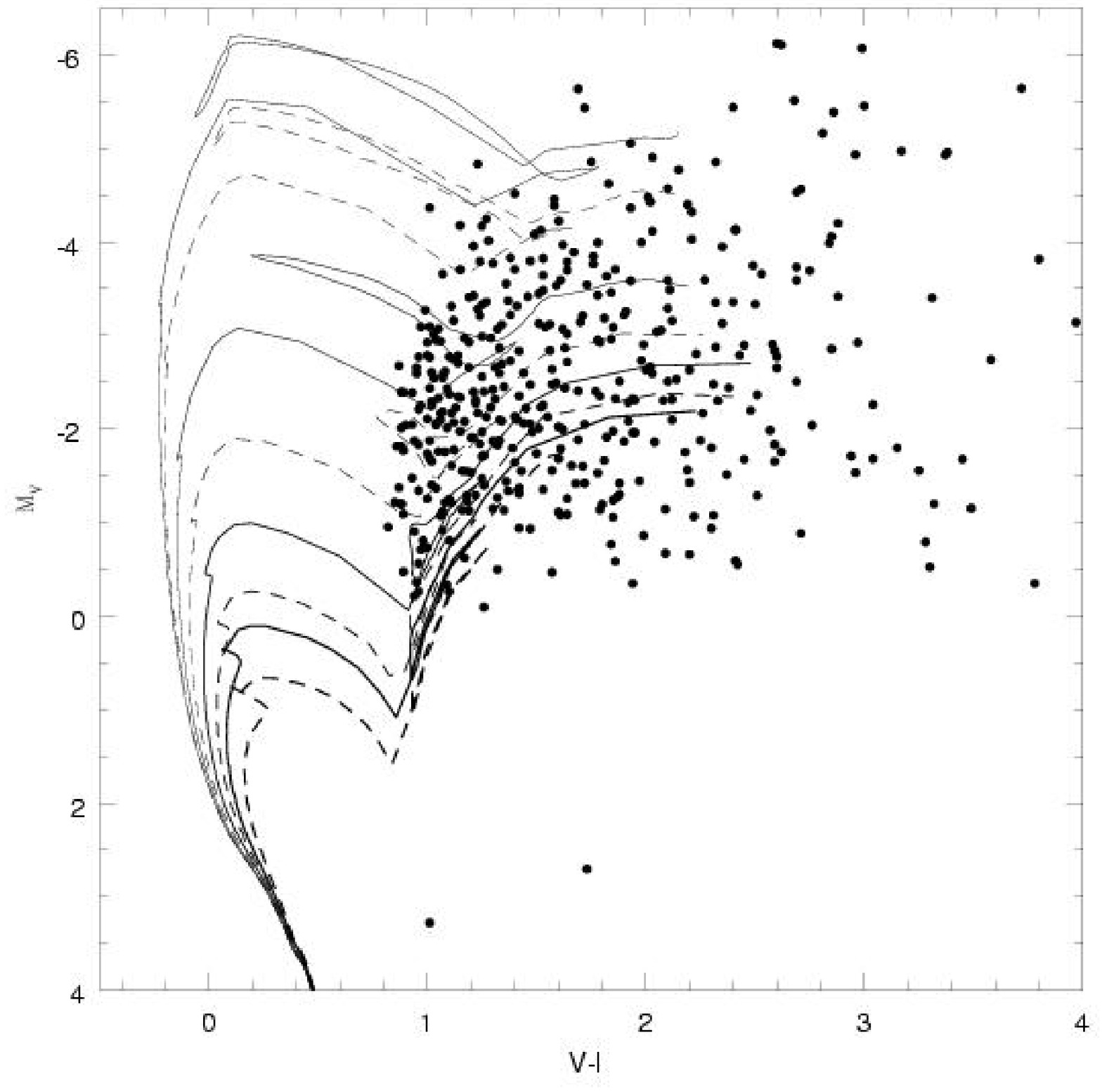}
      \caption{\label{Fig:dhrY}
       HR diagram of group Y. The absolute magnitude used here is
       $M_V$. Isochrones of Lejeune \& Schaerer\ (2001) for $Z=0.008$ and
       log(age(yr))= 7.4, 7.6, 8, 8.3, 8.55, 8.75, 8.85, 9. 
       $V-I$ indices were computed from the colour transformation of Platais et
       al.\ (2003)} 
   \end{figure}

\noindent The first group is one with 'young' kinematics (yellow
group on Fig.~\ref{Fig:UVluri}), with a small velocity dispersion and
with a vertex deviation of $l_v = 16.4^\circ$. Quite remarkably, the value
of $\langle M_{Hp} \rangle$ for this group is 0.68 (see Table~2), 
the brightest among all 6 groups. 
An important property of
this group is thus its young kinematics coupled with its large average
luminosity: the most luminous giants and supergiants \footnote{almost all
  stars {\it a priori} flagged as supergiants (see Sects.~2.2, 3.2 and
  Fig.~5) indeed belong to group Y} in our sample thus appear to have a
small dispersion in the $UV$-plane, in agreement with the general idea that
younger, more massive giants are predominantly found at larger luminosities
in the Hertzsprung-Russell diagram and are, at the same time, concentrated near the origin of
the $UV$-plane. It must be stressed that nothing in the LM method can induce such a
correlation between velocities and luminosity artificially. This result must
thus be considered as a robust result, even more so since the group Y of young
giants was present in all solutions, irrespective of the number of groups
imposed. This group is centered on the usual antisolar motion\ (Dehnen \&
Binney 1998) in $U$ and $W$ as seen in Table~2.

Fig.~\ref{Fig:dhrY} locates stars from group Y in the HR diagram, constructed from the $V-I$ colour index as provided by the colour transformation of Platais et
al. (2003), based on the {\it measured} $H_p - V_{T2}$ colour index (and is thus more accurate than the 
Hipparcos $V-I$ index; see Appendix for more details).
The isochrones of Lejeune \& Schaerer\ (2001), for a typical metallicity $Z=0.008$,
indicate that the age of stars from group Y is on the order of several $10^6$ to a few $10^8$~yr. 

It must be remarked at this point that the observed members of group Y, as displayed on 
Fig.~\ref{Fig:dhrY}, appear to be brighter than the true average absolute magnitude for the 
group listed in Tables~1 and 2 ($\langle M_{Hp} \rangle = 0.68$) and assigned by the LM method. 
This is a natural consequence of the 
Malmquist bias (Malmquist 1936) for a magnitude-limited sample. Nevertheless, even the true average absolute magnitude makes group Y the  brightest and youngest one.

\noindent {\it Group HV: The high-velocity stars}

\begin{figure}
   \centering
   \includegraphics[width=8.5cm]{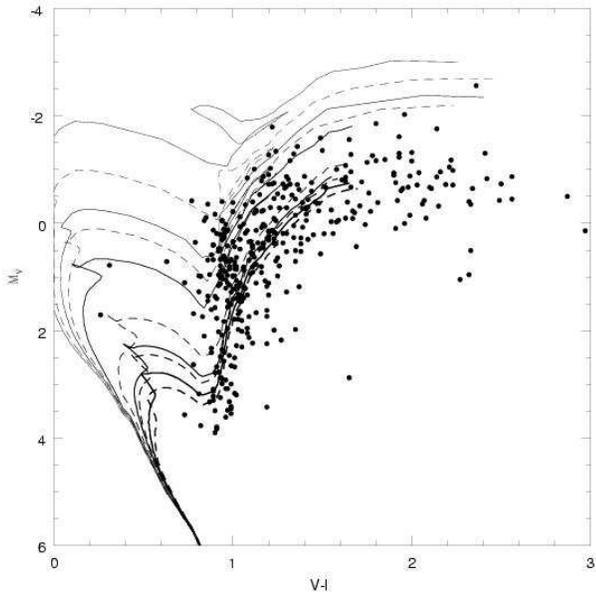}
      \caption{\label{Fig:dhrHV}
       HR diagram of group HV. Isochrones of Lejeune \& Schaerer\ (2001) for
       $Z=0.008$ and log(age(yr))= 8.3, 8.55, 8.75, 8.85, 9, 9.3, 9.45. 9.5, 9.6,
       9.7.  $V-I$ indices were computed from the colour transformation of Platais et
       al.\ (2003)} 
   \end{figure}

This group is composed of high-velocity stars (represented in blue on
Fig.~\ref{Fig:UVluri}). Those stars are probably
mostly halo or thick-disk stars, even though the value of the
scale height of this population is poorly constrained (hence the large difference 
between that parameter derived from the full sample -- Table~2-- and from survey stars only -- 
Table~1) because our sampling distance is too 
small: indeed Fig.~\ref{Fig:LMdistances} shows that the distance of most
stars in our sample is smaller than the scale height of the thick disk ($665 \, {\rm
  pc} < Z_{0,{\rm thick}} < 1000 \, {\rm pc}$). 
  This group represents about 10\% of the whole sample (Table~2, although, like $Z_0$, 
  this parameter is not well constrained), 
  and this value is consistent   
 with the mass fraction of the thick disk relative to the thin disk in the solar
neighbourhood. 

%{\bf The problem of the too small
%sampling distance results in large differences between the model using
%survey stars and the model using the whole sample, even in the fraction of
%stars belonging to the group (see Tables 1 and 2). We choose to rely on the
%results from the whole sample, but refrain from providing too accurate results
%for this group. 

Going back to the different results obtained for the
velocity dispersions in Eqs. (8) and (18), notice that 247 stars (out of
401) of this group have relative errors on their Hipparcos parallax higher
than 20\%.

Clearly we see on
Fig.~\ref{Fig:dhrHV} that stars of this
group are old, mostly older than 1~Gyr and it seems clear that some stars at the
bottom of the HR diagram are as old as the Galaxy itself. It
is also striking that the
envelope of this kinematically 'hot' group seems to correspond to a portion
of circle approximately
centered on a zero-velocity frame  with respect to the galactic center, and
with a radius of the order of $280$~\kms. We refrain,
however, from providing numerical values for the position of this circular
envelope in the $UV$-plane, because of the rather limited number of stars
defining this limit and the absence of antirotating halo stars in our
sample. Note that a
lower limit to the local escape velocity is provided by the 
velocity $(U^2 + (V+220)^2 + W^2)^{1/2} = 281$~\kms\ of
HIP~89298, the fastest star in group HV.

The detection of this group of high-velocity stars
'cleans' the sample and allows us to study the fine structure of the velocity
distribution of disk stars. 

\noindent {\it Group HyPl: Hyades-Pleiades supercluster}

\begin{figure}
   \centering
   \includegraphics[width=8.5cm]{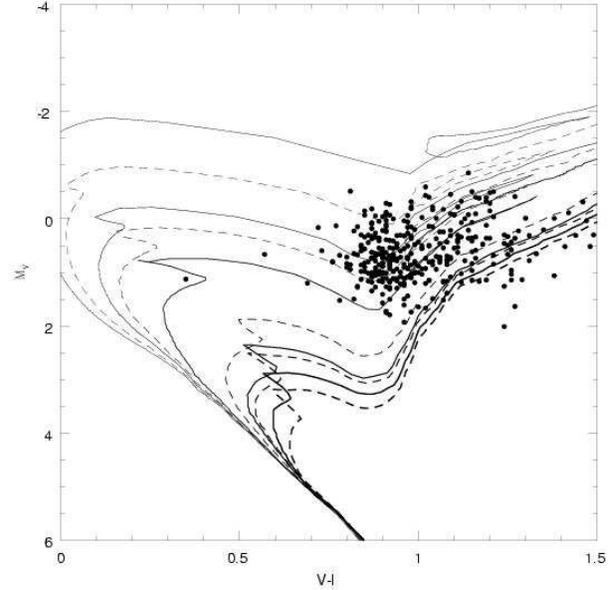}
      \caption{\label{Fig:dhrHyPl}
       HR diagram of group HyPl. Isochrones of Lejeune \& Schaerer\ (2001) for
       $Z=0.02$ and log(age(yr))= 8.3, 8.55, 8.75, 8.85, 9, 9.3, 9.45. 9.5, 9.6,
       9.7. We calculated V-I using the colour transformation of Platais et al.\
       (2003). The stars at the bottom-right of the diagram could be very
       old metal rich stars (in contradiction with the age-metallicity relation
       which is obviously not correct if no other factors are taken into
       account).
       } 
\end{figure}

The group represented in red in Fig.~\ref{Fig:UVluri} occupies the
well-known region of the Hyades and Pleiades superclusters. The large spatial
dispersion of the HyPl stars clearly hints at their supercluster rather than
cluster nature, since they are spread all over the sky with a wide range
of distances (up to 500~pc). The Hyades ($\langle U \rangle =-40$~\kms,
$\langle V \rangle =-25$~\kms, see Dehnen 1998) and
Pleiades ($\langle U \rangle =-15$~\kms, $\langle V \rangle =-25$~\kms, see
Dehnen 1998) superclusters are
known since Eggen\ (1958, 1975). It was also noticed\ (Eggen 1983) that several young
clusters (NGC 2516, IC~2602, $\alpha$ Persei) have the same $V$-component
as those superclusters. We obtain $\langle V \rangle =
-20.3$~\kms for our supercluster structure, and thus refine the old
value $\langle V \rangle =-25$~\kms, with in addition the fact that the group is tilted in
the anti-diagonal direction of the $UV$-plane, i.e., it has a negative
vertex deviation ($l_v = -8.7^\circ$).  The concept of branches in the 
$UV$-plane\ (Skuljan et al. 1999), of
quasi-constant $V$ but slightly tilted in the
anti-diagonal direction, will be further discussed below in relation with the other groups.

Metallicity is available for 17 stars of the HyPl group\ (McWilliam 1990),
and it is interesting to remark that the average metallicity seems to
be close to solar ([Fe/H]$=0$; Fig.~\ref{Fig:met}) as already noticed by Chereul \& Grenon\
(2001). This is quite unusual for giant stars. The metallicity of the stars from group~B
for example is centered on [Fe/H]$= -0.2$. Plotting the HyPl stars in a HR diagram 
(Fig.~\ref{Fig:dhrHyPl}) and using the solar-metallicity isochrones of Lejeune \&
Schaerer\ (2001) reveals that the stars forming the HyPl group are by far not coeval 
(despite the fact that  HyPl stars have a
smaller age spread than the Si and He groups discussed below, especially because the HyPl group is 
lacking young supergiants). 
%The ages are not very precisely known of
%course because we do not know the metallicity of most of 
%the giants present in that group (the metallicities of McWilliam 1990 are
%just an indication). 
Although a precise determination of the ages of HyPl stars would require the knowledge of their 
metallicities, it seems nevertheless clear that there
are a lot of clump stars with ages reaching 1~Gyr, and that there are
clearly stars older than 2~Gyr, in sharp contrast with the ages of about 
80 and 600~Myr for the Pleiades and Hyades clusters themselves. This result of age
heterogeneity, already noticed by Chereul \& Grenon\ (2001) for the Hyades
supercluster, is an important clue to identify the origin of supercluster-like structures, as  discussed in Sect.~3.3.7. 
% A photometric and metallicity analysis of the
%giants of this group could bring very precise results on the age of this
%group and would thus be of great interest.

Finally, the value $\langle W \rangle = -4.8 \pm 0.8$~\kms\ differs significantly from the 
vertical solar motion ($W_\odot = 7$ to 8~\kms; see Sects.~\ref{Sect:mostprecise}, \ref{Sect:MonteCarlo}, and group B in Tables 1 and 2), indicating that the group has a slight net vertical motion.

\noindent {\it Group Si: Sirius moving group}

\begin{figure}
   \centering
   \includegraphics[width=8.5cm]{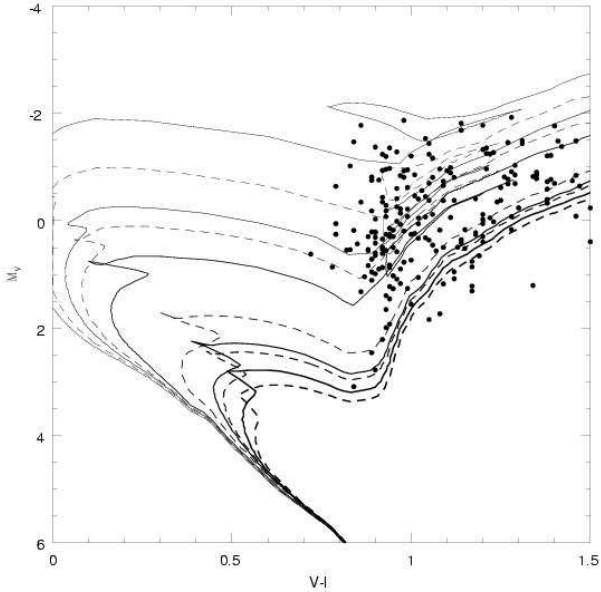}
      \caption{\label{Fig:dhrSi}
       HR diagram of group Si. Isochrones of Lejeune \& Schaerer\ (2001) for
       $Z=0.008$ and log(age(yr))= 8.3, 8.55, 8.75, 8.85, 9, 9.3, 9.45. 9.5, 9.6,
       9.7. } 
   \end{figure}

The Sirius moving group (represented in
magenta in Fig.~\ref{Fig:UVluri}) is known since Eggen\ (1958, 1960) and is
traditionally located at $\langle U \rangle = 10$~\kms , $\langle V \rangle
=-5$~\kms\ (Dehnen 1998). The LM method refines
those values and locates it at $\langle U \rangle =6.5$~\kms, $\langle V
\rangle =3.9$~\kms. The spatial
distribution once again indicates that this group has a supercluster-like
nature. Metallicity from McWilliam (1990) is available for 12 stars of the Si group 
(Fig.~\ref{Fig:met}). Contrarily to the situation prevailing for the 
HyPl group, the metallicity distribution within the Si group appears similar to that for the bulk 
of the giants in the solar neighbourhood, as represented by group~B. 
Isochrones for a typical value of $Z = 0.008$ 
(Fig.~\ref{Fig:dhrSi}) indicate that the ages are widely spread, even more so
than for the HyPl group.  

\noindent {\it Group He: Hercules stream}

\begin{figure}
   \centering
   \includegraphics[width=8.5cm]{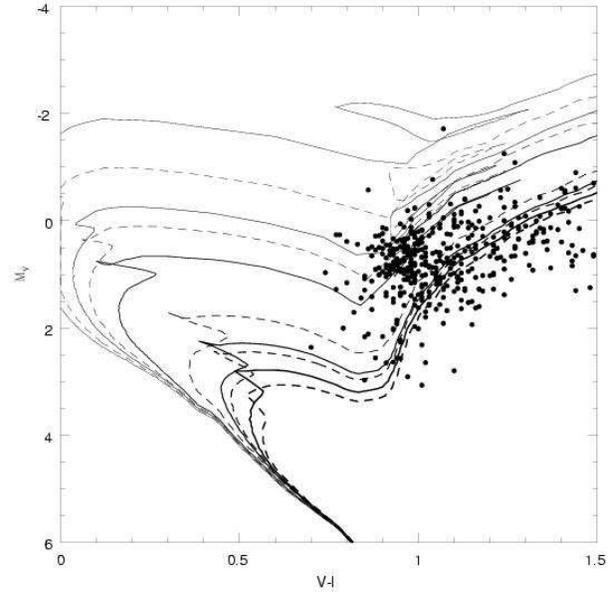}
      \caption{\label{Fig:dhrHe}
        HR diagram of group He. Isochrones of Lejeune \& Schaerer\ (2001) for
       $Z=0.008$ and log(age(yr))= 8.3, 8.55, 8.75, 8.85, 9, 9.3, 9.45. 9.5, 9.6,
       9.7. } 
   \end{figure}

\noindent The 'Hercules stream' (represented in
green in Fig.~\ref{Fig:UVluri}), located by the LM method at $\langle U
\rangle = -42$~\kms,
$\langle V \rangle = -51$~\kms has been named by Raboud et al.\ (1998)
the $u$-anomaly.  It corresponds to a global outward radial motion of the
stars which lag behind the galactic rotation: known since
Blaauw\ (1970), it is traditionally
centered on $\langle U \rangle =-35$~\kms, $\langle V \rangle =-45$~\kms (see Fux 2001). 
It is
strongly believed since several years that its origin is of dynamical nature, as further discussed 
in the next section. Once again, the
Hertzsprung-Russell diagram of Fig.~\ref{Fig:dhrHe} indicates a wide range
of ages in this group (for a typical value of $Z = 0.008$). 

Note that the three  groups HyPl, Si and He could  be interpreted in term
of extended branches crossing the $UV$-plane (Skuljan et al. 1999,
Nordstr\"om et al. 2004): one
is Hercules, another is the combination of Hyades and Pleiades, and the
third one is Sirius. The high dispersion of the streams azimuthal ($V$)
component confirms this view. The branches are somewhat tilted along the
anti-diagonal direction. These properties could be understood in the classical theory of
moving groups as discussed in next section but also in the context of a
dynamical origin for the substructure.

\noindent {\it Group B: Smooth Background}

\begin{figure}
   \centering
   \includegraphics[width=8.5cm]{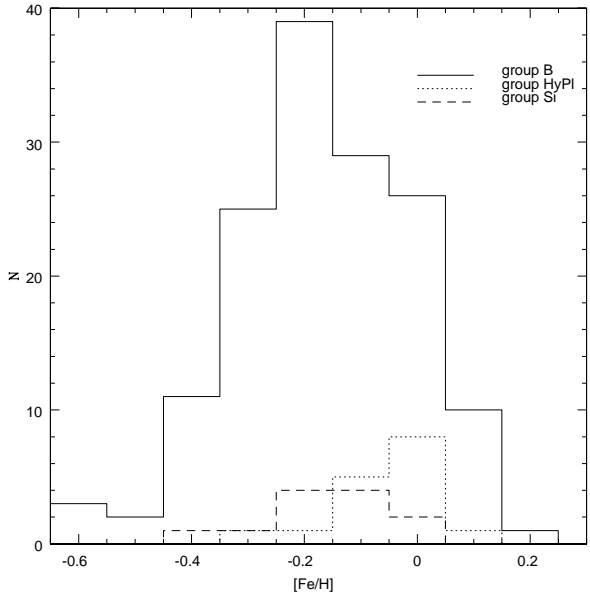}
      \caption{\label{Fig:met}
       Histogram of the metallicity in groups B, HyPl and Si for the stars
       present in the analysis of McWilliam\ (1990)}
   \end{figure}

\begin{figure}
   \centering
   \includegraphics[width=8.5cm]{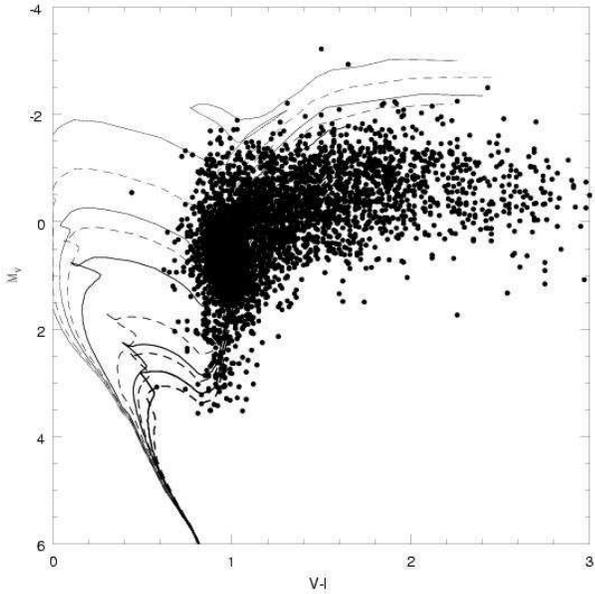}
      \caption{\label{Fig:dhrB}
       HR diagram of group B. Isochrones of Lejeune \& Schaerer\ (2001) for
       $Z=0.008$ and log(age(yr))= 8.3, 8.55, 8.75, 8.85, 9, 9.3, 9.45. 9.5, 9.6,
       9.7. } 
   \end{figure}

\noindent Most stars are part of an 'axisymmetric', 'smooth' background
represented in black in Fig.~\ref{Fig:UVluri}. The
average metallicity of this group seems to be slightly subsolar
(Fig.~\ref{Fig:met}) as expected
for a sample of disk giants: Girardi \& Salaris\ (2001) found an average
metallicity of [Fe/H]=$-0.12 \pm 0.18$, while for their
sample of F and G dwarfs, Nordstr\"om et al.\ (2004) found [Fe/H]=$-0.14 \pm
0.19$. Using
isochrones for a typical value of $Z=0.008$ (see Fig.~\ref{Fig:dhrB}), we
see that there is a large spread in age. This is typical of the mixed population of the galactic
disk, composed of stars born at many different epochs since the birth of
the Galaxy. 

In the $UV$-plane, the velocity ellipsoid of this group is
not centered on the value commonly accepted for the antisolar
motion:  it is centered instead on $\langle U \rangle =
-2.78\pm1.07$~\kms.
However, the full data set (including the various superclusters)
{\it does} yield the usual value for the solar motion (see Sect.~3.2). This
discrepancy clearly raises the essential question of how to derive the
solar motion in the presence of streams in the solar neighbourhood: does
there exist in the solar
neighbourhood a subset of stars having no net radial motion? If the smooth
background is indeed an axisymmetric background with no net radial motion,
we have found a totally different value for the solar motion. Nevertheless,
we have no strong argument to assess that this is the case, especially if
the `superclusters' have a dynamical origin, 
as proposed in 
the next section. Moreover, the group Y of young giants is centered on the commonly
accepted value of $\langle U \rangle =-10.41\pm0.94$~\kms\ and this difference between groups~Y
and B clearly prevents us from deriving without ambiguity the solar
motion. If the giant molecular clouds (GMC's) from which the young stars
arose are on a circular orbit, then the $\langle U \rangle$ value of group~Y 
is the acceptable one for the
antisolar motion, but nothing proves that the GMC's are not locally moving
outward in the Galaxy under the effect of the spiral pattern. No value can
thus at the present time be given for the radial solar motion but only
some different estimates depending on the theoretical hypothesis we make on
the nature of the substructures observed in velocity space. Theoretical investigations
and dynamical simulations thus appear to be the only ways to solve this problem. On
the other hand, the value of $\langle W \rangle = -8.26\pm0.38$~\kms\ 
for group~B is in accordance with the usual motion of the sun perpendicular
to the galactic disk (see Sect.~3.2), and thus seems to be a reliable value
because streams have a smaller effect on the vertical motion of stars.

\subsubsection{Physical interpretation of the groups: dynamical streams?}

Several mechanisms may be responsible for the substructrure observed in
velocity space in the solar neighbourhood. Hereafter, we list them and
confront them with the results of our kinematic study.

A first class of mechanisms is that associated with
inhomogeneous star formation
responsible for a deviation from equilibrium in the solar neighbourhood:
this theory states that a large number of stars are formed (almost)
simultaneously in a certain region of the Galaxy and create a
cluster-like structure with a well-defined position and velocity. After
several galactic rotations, the cluster will evaporate and form a tube called
'supercluster'. Stars in the 'supercluster' still share common $V$
velocities when located in the same region of the tube (for example in the
solar neighbourhood) for the following reason, first put forward by Woolley\
(1961): if the present galactocentric radius of a star on a quasi-circular
epicyclic orbit equals that of the sun (denoted $\varpi_\odot$), and if
such a star is observed with a peculiar velocity $v = V + V_\odot$, then
its guiding-center radius $\varpi_g$ writes 

\begin{equation}
\label{Eq:epicyclic}
\varpi_g = \varpi_\odot - x_g = \varpi_\odot + \frac{v}{2 B}
\end{equation}
where $x_g$ is the position of its guiding-center in the cartesian
reference frame associated with the LSR (in the solar neighbourhood
approximation, the impact of $y_g$ is negligible), and $B$ is the second Oort
constant. Woolley\ (1961) pointed out that disk stars (most of which move
on quasi-circular epicyclic orbits) which formed at the same place and
time, and which stayed together in the Galaxy after a few galactic
rotations (since they are all currently
observed in the solar neighbourhood) must necessarily have the same period
of revolution around the Galactic center, and thus the same guiding-center
$\varpi_g$, and thus the same velocity $V = v - V_\odot$ according to
Eq. (\ref{Eq:epicyclic}). This theory has
thus the great advantage of predicting extended horizontal branches
crossing the $UV$-plane (similar to those observed in our sample on
Fig.~\ref{Fig:UVluri} and already identified by Skuljan et
al. 1999, and by Nordstr\"om et al. 2004 in their sample of F and G dwarfs). Moreover, it is easy to understand in this framework that the
Group HyPl seems to be more metal-rich than the smooth background (see
Fig.~\ref{Fig:met}). However, this theory does not explain the tilt of the
branches that we observe in our sample. Moreover, to explain the wide range
of ages observed in those branches (see Figs. \ref{Fig:dhrHyPl},
\ref{Fig:dhrSi}, and also Chereul \& Grenon 2001), the stars must
have formed at different epochs (see e.g. Weidemann et al. 1992) out of
only two large molecular clouds (one 
associated with the Sirius branch and another with the Hyades-Pleiades
branch). Chereul et al.\ (1998) suggested that the supercluster-like velocity
structure is just a chance juxtaposition of several cluster remnants, but
this hypothesis requires extraordinary long
survival times for the oldest clusters (with ages $> 2$ Gyr) in the supercluster-like structure. This long survival time is made unlikely by the
argument of Boutloukos \& Lamers\ (2003) who found that clusters within 1
kpc from the Sun having a mass of $m \times 10^4 \, {\rm M}_\odot$ can survive
up to $m$ Gyr in the Galaxy: indeed heavy clusters of more than $2 \times
10^4 \, {\rm M}_\odot$ are probably very rare in the {\it disk}. 

An explanation for the wide range of ages observed in the superclusters
could be the capture of some older stars by the high concentration of mass
in a molecular cloud, at the time of the formation of a new group of stars
(while other stars would be scattered by the molecular cloud) . Though this
hypothesis could be tested in $N$-body simulations involving gas, it would
imply that the local clumps of the potential not only perturb the
motion of stars but dominate it, which is in contradiction with the high
predominance of the global potential on the local clumps. This explanation is thus highly unlikely too. 

The second class of mechanisms involves the theory of hierarchic formation
of the galaxies: following this theory, galaxies were built up by the
merging of smaller precursor structures. It is known since the discovery of the
absorption of the Sagittarius dwarf galaxy by the Milky Way\ (Ibata et
al. 1994) that some streams in the Galaxy are remnants of a merger with a
satellite galaxy. Helmi et al.\ (1999) showed that some debris streams are also
present in the galactic halo near the position of the Sun. Such streams
could also be present in the velocity substructure of the disk: it
seems to be the case of the Arcturus group at $U \simeq 0$~\kms, $V \simeq
-115$~\kms, as recently proposed by Navarro et al.\ (2004). In this
scenario, the streams observed in our kinematic study could also be the
remnants of merger events between our Galaxy and a satellite galaxy. The
merger would have triggered star formation whereas the oldest giant stars
would be stars accreted from the companion galaxy. A merger with a
satellite galaxy would moreover induce a perturbation in $W$, as we observe
in group HyPl in which 
$\langle W \rangle = -4.8 \pm 0.8$~\kms. If the hierarchical ('bottom-up') cosmological
model is correct, the Milky Way system should have accreted and subsequently tidally destroyed approximately 100 low-mass
galaxies in the past 12 Gyr (see Bullock \& Johnston 2004),  which leads to
one merger every 120 Myr, but the chance that two of them
(leading to the Hyades-Pleiades and Sirius superclusters) have left such
important signatures in the disk near the position of the Sun is
statistically unlikely (although  not impossible). 

The third class of mechanisms is the class of purely dynamical ones. A
dynamical mechanism that could cause substructure in the local
velocity distribution is the disturbing
effect of a non-axisymmetric component of the gravitational
potential, like the rotating galactic bar. The Hercules stream was
recently identified with the bimodal character of the local velocity
distribution\ (Dehnen 1999, 2000) due to the rotation of the bar if the Sun
is located at the bar's outer Lindblad resonance (OLR). Indeed, stars in the Galaxy
will have their orbits elongated  along or perpendicular to the major
axis of the bar (orbits respectively called the LSR and 
OLR modes in the terminology of Dehnen 2000),  depending upon their position relative to the resonances, and
both types of orbits coexist at the OLR radius. Moreover, all orbits are
regular in a 2-dimensional (2D) axisymmetric potential, but the perturbation of the triaxial bar will induce some chaos. Fux\ (2001) showed that in the region of the Hercules stream in velocity space, the chaotic
regions, decoupled from the regular regions, are more
heavily crowded. The Hercules stream has thus also
been interpreted as an overdensity of chaotic orbits\ (Fux 2001) due to the
rotating bar. Quillen\ (2003) has confirmed that when the effect of
the spiral structure is added to that of the bar, the 'chaotic' Hercules stream remains a strong feature of the local DF, whose boundaries are refined by
the spiral structure. This stream in the local DF seems thus related
to a non-axisymmetric perturbation rather than to a deviation from
equilibrium due to inhomogeneous star formation. Following
M\"uhlbauer \& Dehnen\ (2003), the Galactic bar naturally
induces a non-zero vertex deviation on the order of $10^\circ$ and the
vertex deviation found in Sects. 3.1 and 3.2 would thus be partly
related to the Hercules stream. Nevertheless, the other streams present in
the data are also partly responsible for the vertex deviation. The likely
dynamical origin of the Hercules stream has been the first example of a
non-axisymmetric origin for a stream in velocity space: the other streams
could thus be related to other non-axisymmetric effects. Chereul \& Grenon\
(2001) proposed that the Hyades supercluster represents in fact an extension
of the Hercules stream, but the LM method has shown that the two features are
clearly separated in the $UV$-plane. 

De Simone et al.\ (2004) have shown that the structure of the local DF
could be due to a lumpy potential related to the presence of transient spiral waves in the shearing sheet (i.e. a small portion of an
infinitesimally thin disk that can be associated with the solar
neighbourhood). Julian \& Toomre\ (1966) have studied, in the shearing sheet,
the response of a stellar disk to a point-like
perturbation of the potential. They found that this response is in the form
of density waves whose wavecrests swing around from leading to
trailing. These waves are transient because their amplitudes are amplified
transitorily and then die away. The duration of these transient spiral
waves is of the order of an orbital period,
i.e. of the order of  $10^8$ years, and they impart
momentum to stars: this mechanism tends to put stars in some
specific regions of the $UV$-plane in the simulations of De Simone et al.\
(2004), thus creating streams as observed in our sample. These spiral waves cause radial migration in the galactic
disk near their corotation radius, while not increasing the random motions
and preserving the overall angular
momentum distribution\ (Sellwood \& Binney 2002). The seemingly
peculiar chemical composition
of the group HyPl (i.e., a metallicity higher than average for field giants, as suggested 
by the  17 giant stars analyzed  by  McWilliam 1990 and displayed in Fig.~\ref{Fig:met}, 
%{\bf which is too
%small a sample to give sufficient evidence}, but also
also reported by Chereul \& Grenon 2001) thus suggests that the group has a 
common galactocentric origin in the inner Galaxy (where the interstellar
medium is more metal-rich than in the solar neighbourhood) and that it was
perturbed by a spiral wave at a certain moment. This specific  scenario (Pont
et al., in preparation) would
explain why this group is composed of stars sharing a common metallicity
but not a common age. The group Si could also be a clump recently formed by the
passage of a transient spiral. 

The simulations of De Simone et al.\ (2004)
can create streams with a {\it range of 3 Gyr or more in age}, and this is
thus a mechanism that can explain that main result of our study. Another 
characteristic of the simulations is that they tend to reproduce the
observed branches, and their origin ultimately lies in the same mechanism
as that elucidated by Woolley\ (1961). However, the tilt of the branches in
the $UV$-plane is not reproduced by the simulations. 

A question which arises in the framework of this dynamical scenario is the
following: is it by chance that such a large number of young clusters and associations are 
situated on the same branches in the $UV$-plane (more than
half of the OB associations listed in Table A.1 of de Zeeuw 1999 are
situated in the regions of the HyPl and Si groups on
Fig.~\ref{Fig:UVluri})? Probably not, which suggests that the same
transient spiral which gave their peculiar velocity to the HyPl and Si
groups has put these young clusters and associations along the same $UV$
branches. For
example, it is 
quite striking that the Hyades cluster itself is metal-rich with a mean
[Fe/H]= 0.13\ (Boesgaard 1989), thus pointing towards the same   galactocentric origin 
as  the  HyPl group. The entire cluster could have been shifted in radius
while remaining bound since the effect of a spiral wave on stars
depends on the stars' phase with respect to the spiral, and the phase
does not vary much across the cluster. Moreover, since
most clusters and associations are young, they should not have crossed many
transients, which suggests that one
and the same spiral transient could have formed some clusters and
associations (by boosting
star formation in the gas cloud) and could at the same time  have given them their peculiar velocity in the $UV$-plane. The relation
between spiral waves and star formation is indeed well established
(e.g. Hernandez et al. 2000 who found a star formation rate $SFR(t)$ with an
oscillatory component of period 500 Myr related to the spiral pattern). As
a corollary, we conclude that the dynamical streams observed among K and M
giants are {\it young} kinematic features: integrating backwards (in a
smooth stationnary axisymmetric potential) the orbits of the stars
belonging to the streams makes thus absolutely no sense, and
reconstructing the history of the local disk from the present data of stars
in the solar neighbourhood becomes tricky. In this dynamical
scenario, the deviation from dynamical equilibrium that is present among
samples of young stars is closely related to the deviation from axisymmetry
existing in the Galaxy. Of course, we do not exclude that the position
  of some clusters and OB associations in the same region of velocity space
  as the dynamical streams could be the result of chance\ (Chereul et
  al. 1998).

If this dynamical scenario is correct, the term 'dynamical stream' for the branches in
velocity space seems more appropriate than the term 'supercluster' since
they are not caused by contemporaneous star formation but rather involve stars
that do not share a common place of birth: stars in the streams just share at
present time a common velocity vector.

It should be noted that those non-axisymmetric perturbations, as well as
the minor mergers, could lead to some asymmetries in the spatial
distribution of stars in the galactic disk on a
large scale (see Parker et al. 2004). Since our sample does not cover the
whole sky, and is anyway restricted to the solar neighbourhood, we are not
in a position to detect those asymmetries.

To conclude this section, let us stress that the dynamical,
non-axisymmetric theory to explain the substructure observed in the
velocity space is largely preferred over the theory which views it as
remnants of clusters of stars sharing a common
initial origin, essentially because of the wide range of ages of the stars
composing the identified subgroups. In fact, we stress that the deviation
from equilibrium among young stars\ (Dehnen \& Binney 1998) and the
dynamical streams exhibited in this paper
could be closely related to each other. The presence of dynamical streams in
our sample of K and M giants is clearly responsible for the vertex
deviation found for late-type giants, but we even suggest that the same
origin could hold as well for early-type stars. It is indeed quite striking on
Fig.~\ref{Fig:UVluri} that {\it the group Y extends just far enough in the
  $UV$-plane to touch both the HyPl and Si branches}. The presence of stars
from these two streams in group Y (sent in that region of the
$UV$-plane at the time of their formation because of the peculiar velocity
imparted by the spiral wave that created them) imposes a very specific value to the
vertex deviation (ranging from $15^\circ$ to $30^\circ$ and more), in
agreement with the high values often observed for young stars (see Dehnen
\& Binney 1998). This idea
that the vertex deviation for younger populations could in fact have the
same {\it dynamical} origin as the vertex deviation for old ones was
already proposed by Mayor\ (1972, 1974). Here, we also argue that even the
specific initial conditions of young groups of stars could be due to the
same phenomenon.  

However, nature is of course not so simple and the features of the
DF are presumably related to a mixture of several phenomena. Notably, the
initial conditions in the simulations should be more complex than the
simple 2D Schwarzschild velocity ellipsoid used by De Simone et al.\ (2004): in fact, purely {\it axisymmetric} substructure could already be present in the solar
neighbourhood (see the structure of the
$UV$-plane in Dejonghe \& Van Caelenberg 1999). Other phenomena that could
have an influence on the structure of velocity space are the following: a
triaxial or clumpy dark halo, giant molecular clouds, and close encounters with
the Magellanic Clouds (Rocha-Pinto et al. 2000). Theoretical
investigations in this area should thus clearly be pursued.

\section{Summary and perspectives}

This paper presented the first kinematic analysis of
5311 K and 719 M giants (after excluding the binaries for which no
center-of-mass velocity could be estimated) in the solar neighbourhood
which included radial velocity data from an important survey performed with
the CORAVEL spectrovelocimeter. We also used proper motions from the
Tycho-2 catalogue.  

First, we analyzed the kinematics of the sample restricted to the 2774
stars with parallaxes accurate to better than 20\%, and then we made full
use of the 6030 available stars and evaluated the kinematic parameters with
a Monte-Carlo method. We found that the asymmetric drift is larger for
M giants than for K giants due to the fact that the M giants must be a little
older than the K giants on average. We also found the usual value for the
solar motion when assuming that the whole sample has no net radial and
vertical motion (Sect. 3.2).   

Then a maximum-likelihood method, based on a bayesian approach (Luri et
al. 1996, LM method), has been applied to the data and allowed us to derive
simultaneously maximum likelihood estimators of luminosity and kinematic
parameters, and to identify subgroups present in the sample. 
Several subgroups can be identified with
known kinematic features of the solar neighbourhood (namely the
Hyades-Pleiades supercluster, the Sirius moving group and the Hercules stream). Isochrones in the
Hertzsprung-Russell diagram reveal a very wide range of ages for stars
belonging to these subgroups. We thus conclude that this substructure in
velocity space is probably related to the dynamical perturbation by
transient spiral waves (as recently modelled by De Simone et
al. 2004) rather than to cluster remnants. The emerging picture is thus one of dynamical streams pervading
the solar neighbourhood. A possible explanation for the presence of young group/clusters in the same area of the
$UV$-plane as those streams is that they have been put there by the spiral wave associated
with their formation, while kinematics of the older
stars of our sample have also been disturbed by the same wave. The seemingly
peculiar chemical composition of the Hyades-Pleiades stream (also reported
by Chereul \& Grenon 2001) suggests that this stream originates  
from a specific  galactocentric  distance and that it was
perturbed by a spiral wave at a certain moment and radially pushed by the
wave in the
solar neighbourhood\ (see Sellwood \& Binney 2002; Pont et al. in
preparation). This would explain why this stream is composed of stars sharing a common metallicity but not a common
age. A careful metallicity analysis of this stream would be of great
interest in order to confirm this scenario. The Sirius moving group could also be a feature recently formed by the
passage of a transient spiral, while the Hercules stream would be related to
the bar's outer Lindblad resonance\ (Dehnen 1999, 2000, Fux 2001). The
position of all these streams in the $UV$-plane is responsible for the
vertex deviation of $16.2^\circ \pm 5.6^\circ$ for the whole sample (see
Sect. 3.2). We
even argue that the vertex 
deviation observed among large samples of early-type stars (see Dehnen \&
Binney 1998) and the
specific kinematic initial conditions of some young open clusters and OB associations
could in fact have
the same dynamical origin as those streams of giants. A better understanding
of the streams should start with a chemical analysis of the stars
belonging to them. As a first step, their photometric indices could be
investigated. Note also that an important consequence of the dynamical
origin of the streams is that it makes no sense to integrate backwards (in a
smooth stationary axisymmetric potential) the orbits of the stars belonging
to them.   

The group of background stars (group B) has a distribution in velocity space close to a 
Schwarzschild
ellipsoid but is not centered on the classical value found for
$-U_\odot$ in Sect.~3.2 when considering the full sample, including the streams 
Instead we find $\langle U \rangle =
-2.78\pm1.07$~\kms.  This discrepancy 
%with the
%value found for the whole set of stars including the various streams
clearly raises the essential question of how to derive the solar
motion in the presence of dynamical perturbations altering the kinematics
of the solar neighbourhood (the net radial motion of stars in the solar
neighbourhood can be of the order of 10~\kms\ in the simulation of De Simone
et al. 2004): does there exist in the solar neighbourhood a subset of stars
having no net radial motion which can be used as a reference against which
to measure the solar motion? 

Theoretical investigations in this area should thus clearly be pursued, and
in particular {\it dynamical modeling}. We have shown that the fine
structure of phase space in the solar neighbourhood cannot be interpreted
in terms of an axisymmetric steady-state model. Nevertheless, an
axisymmetric model revealing  all the fine structure of the axisymmetric distribution function in the solar neighbourhood\ (Dejonghe \&
Van Caelenberg 1999, Famaey et al. 2002, Famaey \& Dejonghe 2003) would
give ideal initial conditions (more complex than a simple 2D Schwarzschild
velocity ellipsoid; see especially the structure of the $UV$-plane in
Dejonghe \& Van Caelenberg 1999) for 3D $N$-body simulations that could
afterwards reproduce some non-axisymmetric features observed in the solar
neighbourhood\ (see Fux 1997, De Simone et al. 2004 for 2D simulations).   

\begin{acknowledgements}

We thank R. Griffin for allowing us to derive the center-of-mass velocity
of binaries from his unpublished CORAVEL data. Most of the existing orbits
for K giants quoted in Table A.1 were in fact derived by R. Griffin,
and we want to recognize here his enormous and valuable contribution to
this field. We also thank G. Traversa and B. Pernier for
their important contribution to the observations with the CORAVEL
spectrometer. This research was made possible by the substantial
financial support received from the {\it Fonds National Suisse pour la
  Recherche scientifique} which funded the development of CORAVEL, and the
operation of the Swiss 1-m telescope at the {\it Observatoire de Haute
  Provence.} 

\end{acknowledgements}

\appendix
\section{Contents of the data table}

Table A.1\footnote{ Table A.1 is only available in
electronic form at the CDS via anonymous ftp to cdsarc.u-strasbg.fr
(130.79.128.5) or via http://cdsweb.u-strasbg.fr/cgi-bin/qcat?J/A+A/ /. Nevertheless, the first page of Table A.1 is printed hereafter.}
contains 6691 lines, and contains the following information in the
successive columns (note that missing data are replaced by null values):\\
\begin{itemize}  
\item[1] HIP number.
\item[2] HD number.
\item[3-4] BD number, only when there is no HD number.
\item[5-6] Right ascension and declination in decimal degrees from fields
  H8 and H9 of the Hipparcos Catalogue (ICRS, equinox 2000.0; epoch 1991.25).
\item[7-8] Hipparcos parallax and standard error. 
\item[9-10] $\mu_{\alpha*} \equiv  \mu_\alpha \cos \delta$ from Tycho-2
and standard error.
\item[9-10] $\mu_\delta$ from Tycho-2
and standard error. For 291 stars, null values
are found in columns 9 to 12; the kinematic study made then use of the
Hipparcos proper motions instead of the Tycho-2 ones. In most  cases, the 
absence of a Tycho-2 proper motion
is caused by the fact that the star has a close visual companion or is too
bright for the Tycho detection. 
\item[13-16] Same as columns 9 to 12 for Hipparcos proper motions.
\item[17] The normalized absolute difference between the Hipparcos and
Tycho-2 proper motions:
$\Delta \mu = |\mu_{\rm Hip} - \mu_{\rm Tyc2}|/\epsilon_{\mu}$, where
$\mu = (\mu^2_{\alpha} \cos^2 \delta + \mu^2_\delta)^{1/2}$ and
$\epsilon_{\mu} = (\epsilon_{\mu_{\rm Hip}}^2 + \epsilon_{\mu_{\rm
Tyc-2}}^2)^{1/2}$, where $\epsilon_i$ denotes the standard error of 
quantity $i$.
The quantity $\Delta\mu$ may be used as a diagnostic tool to identify 
long-period binaries (Kaplan \& Makarov 2003). The basic idea behind this tool is the
following. For binary stars with orbital periods much longer than the
duration of the Hipparcos mission, the proper motion recorded
by Hipparcos is in fact the vector addition of the true proper
motion and of the orbital motion. This orbital motion averages out in the 
Tycho-2 value, since it is derived from measurements spanning a much
longer time base (on the order of a century, as compared to 3~yr for
Hipparcos). Therefore, a difference between the Tycho-2 and Hipparcos
proper motions (beyond the combined  error bar encapsulated by 
$\epsilon_{\mu}$) very likely hints at the binary nature of the star.
This diagnostic will be fully exploited in a forthcoming paper devoted
to the binary stars present in our sample. Note that
Fig.~1 presents the histogram of $\Delta\mu$, which
indeed reveals the presence of an abnormally large tail at $\Delta\mu
\ge 1.5$.
\item[18-19] Hipparcos $H_p$ magnitude and associated standard error.
\item[20] Tycho-2 $V_{T2}$ magnitude.
\item[21] $H_p - V_{T2}.$ For visual binaries (flag 4 in column 24), this
colour index has not been listed and the value 0.0 is given instead,
because the $H_p$ magnitude appears to be a composite value for the two
visual components. In most of these cases, the $V_{T2}$ magnitude of the
visual companion is given in the last column of the table. For large amplitude variable stars, 
the $H_p - V_{T2}$ index is meaningless as well, and as been set to 0.  
\item[22] $V-I$ colour index in Cousins' system as provided by field H40
of the Hipparcos catalogue. This index is useful for constructing the
Hertzsprung-Russell (HR) diagram of the sample. It should be stressed that
the H40 field of the Hipparcos catalogue does not provide a directly
measured quantity. It has instead been computed from various colour
transformations based on the $B-V$ index from field H37, which neither is
a directly measured quantity. 
\item[23] $V-I$ colour index in Cousins' system as provided by the colour
transformation based on the {\it measured} $H_p - V_{T2}$ colour index\ (Platais et al. 2003). This value is thus in principle more reliable 
than the Hipparcos H40 value (note, however, that the colour transformation
provided by Platais et al. (2003) had to be extrapolated somewhat, since it
is provided in the range $-2.5
\le H_p - V_{T2} \le -0.20$, whereas our data set goes up to $H_p - V_{T2} =
0.1$). It has been used to draw the HR diagram of the sample.
For large-amplitude variable stars, the median $V-I$ index has been taken
directly from Platais et al. (2003), instead of being computed from the colour transformation based on 
$H_p - V_{T2}$ (set to 0. in those cases).
\item[24] Binarity flag:\\
{}*: no evidence for radial-velocity variations;\\
0: spectroscopic binary (SB), with no orbit available. The star had to be
discarded from the kinematic analysis;\\
1: SB with an orbit available, or with a center-of-mass velocity which
can be estimated reasonably well. Column 25 then contains the system's
center-of-mass velocity. The reference with the orbit used to derive the
center-of-mass velocity is listed in the last column. If no reference
is given, the center-of-mass velocity has been estimated from the
available CORAVEL data;\\  
2: supergiant star, with a substantial
radial-velocity jitter (see Fig.~5);\\ 3: uncertain
case: either SB or supergiant;\\ 4: visual binary with a companion less
than $6''$ away (as listed by the Tycho-2 catalogue);\\ 
5,6,7,8: as
0,1,2,3 but for a visual binary. It should be stressed here that visual
binaries have not been searched for exhaustively among our target stars. 
Whenever the Tycho-2 catalogue lists a companion star less  than $6''$ 
away from the target star, the binarity flag has been set to 4. In most
of these cases, the Hipparcos
$H_p$ magnitude corresponds to the composite magnitude and the
Tycho-2 proper motions are identical for the two components. The
$V_{\rm T2}$ magnitude of the companion is listed in the last column of
the table;\\ 9: binary supergiant with an orbit available. 
\item[25-26] Average radial velocity (based on CORAVEL observations) or
center-of-mass velocity for SBs (the last column provides the
bibliographic code of the reference providing the orbit used), and
standard error (set to 0.3~km~s$^{-1}$ in the case of center-of-mass
velocity).
\item[27] The absolute $H_p$ magnitude, corrected for interstellar
reddening (according to the model of Arenou et al. 1992), and based on
the LM distance listed in column 28.
\item[28-29] The maximum-likelihood distance based on the LM estimator
(see Sect.~\ref{Sect:LM}), and its associated standard error (see Eq.~(\ref{Eq:errdist})).
\item[30] The interstellar absorption $A_V$, based on the LM distance
and the model of Arenou et al. (1992).  
\item[31-33] The $U, V$ and $W$ components of the heliocentric space
velocity deduced from the LM method (corrected for the galactic
differential rotation to first order using $A = 14.82
  \,{\rm km}\,{\rm s}^{-1}{\rm kpc}^{-1}$ and $B = -12.37 \,{\rm km}\,{\rm
  s}^{-1}{\rm kpc}^{-1}$).  
\item[34-35] The most likely group to which the star belongs, and the
associated probability. The various groups are the following:\\
1 or Y:  stars with a young kinematics;\\
2 or HV: high-velocity stars;\\ 
3 or B: stars defining the smooth background (the most populated group);\\ 
4 or HyPl: stars belonging to the Hyades-Pleiades stream;\\
5 or Si: stars belonging to the Sirius stream;\\
6 or He : stars belonging to the Hercules stream.
\item[36] In the case of a spectroscopic binary with an available orbit: 
Bibliographic code (according to the standard ADS/CDS coding) of
the reference listing the orbit used as the source of the center-of-mass
velocity (The code 2005A\&A...???..???J refers to the forthcoming paper by
Jorissen et al. 2005 devoted to the analysis of the binary content of
the present sample). In the case of a visual
binary, the
$V_{\rm T2}$ magnitude of the companion. An asterisk in that column
indicates the presence of a note in the remark file. 
\end{itemize}

It must be stressed that columns 27 to 35 contain model-dependent data,
as they were derived by the LM method. They depend upon, {\it e.g.,} the
particular choice for the values of the Oort constants, the interstellar
extinction model, the {\it a priori} choice of the various distribution
functions,...

\tabcolsep 4pt
\begin{table*}
\caption[]{
The left-hand side of the first page of Table A.1, fully available in electronic form at the CDS. See text for a description of the column 
contents.
}
\begin{tabular}{rrrrrrrrrrrrrrrrrrrrrrrrrrrrrrrrrrrrrrrrrrrr}
\hline
HIP  &   \multicolumn{1}{c}{HD} & \multicolumn{2}{c}{BD}& \multicolumn{1}{c}{$\alpha$}      &    \multicolumn{1}{c}{$\delta$}  &  
\multicolumn{2}{c}{$\pi\;\pm$}  & \multicolumn{2}{c}{$\mu_{\alpha*,T-2} \;\pm$} & \multicolumn{2}{c}{$\mu_{\delta,T-2}\;\pm$} 
&\multicolumn{2}{c}{$\mu_{\alpha*,HIP} \pm$} & \multicolumn{2}{c}{$\mu_{\delta,HIP}\; \pm$} &  
$\Delta\mu$  \cr
     &     &     &           & \multicolumn{1}{c}{($^\circ$)}   &    \multicolumn{1}{c}{($^\circ$)} & \multicolumn{2}{c}{(mas)}    & 
\multicolumn{2}{c}{(mas)} &  \multicolumn{2}{c}{(mas)} &  \multicolumn{2}{c}{(mas)}    & \multicolumn{2}{c}{(mas)} \cr
\hline
    21&224724&  0&    0&   0.06623569&  8.00723437&    5.84& 0.95&    62.5& 1.2&    -0.4& 1.3&   61.89& 0.84  & -0.22& 0.56&  -0.30&\cr
    31&224760&  0&    0&   0.09809390&  2.67547768&    1.84& 1.05&    -2.1& 1.0&     1.3& 1.1&   -4.88& 1.49  & -0.20& 0.69&   1.09&\cr
    36&224759&  0&    0&   0.10327550& 12.26709303&    6.30& 0.96&    49.5& 1.1&    15.6& 1.1&   49.71& 0.85  & 14.27& 0.60&  -0.10&\cr
    73&224826&  0&    0&   0.22045719& 66.84796687&    3.97& 0.69&   -23.4& 1.3&    18.0& 1.3&  -20.78& 0.67  & 18.19& 0.55&  -0.94&\cr
   119&224894&  0&    0&   0.39020229& 44.67544581&    7.49& 0.83&   -11.6& 0.9&   -17.0& 1.0&  -10.61& 0.60  &-18.37& 0.51&   0.41&\cr
   121&224895&  0&    0&   0.39647681& 28.42377416&    3.66& 0.83&   -18.7& 1.1&   -33.6& 1.1&  -18.88& 0.66  &-34.10& 0.46&   0.30&\cr
   123&224891&  0&    0&   0.40024110& 72.23660619&    5.07& 0.70&   -44.5& 0.8&     3.6& 0.9&  -44.73& 0.57  &  3.74& 0.53&   0.17&\cr
   136&224907&  0&    0&   0.42864226& 24.25318080&    5.84& 0.78&    18.1& 0.8&   -29.0& 0.8&   18.50& 0.71  &-28.01& 0.47&  -0.44&\cr
   142&224918&  0&    0&   0.45374866& 66.30600204&   16.04& 0.69&     5.6& 1.2&    13.1& 1.3&    4.74& 0.65  & 12.31& 0.53&  -0.54&\cr
   181&224980&  0&    0&   0.57381667& 60.70319460&    2.51& 0.64&    -7.9& 1.2&   -12.0& 1.3&   -8.91& 0.50  &-11.05& 0.52&  -0.09&\cr
   258&225073&  0&    0&   0.80532807& 17.55239698&    6.28& 0.78&    57.8& 0.9&   -52.8& 0.9&   58.45& 0.74  &-52.87& 0.53&   0.34&\cr
   282&225106&  0&    0&   0.87642718& 19.38157170&    1.62& 0.97&    -0.3& 1.0&   -29.4& 1.0&    0.04& 0.77  &-31.13& 0.59&   1.01&\cr
   288&225105&  0&    0&   0.88875927& 42.35221969&    5.49& 0.97&    99.1& 1.1&   -46.7& 1.1&   98.95& 0.66  &-46.75& 0.61&  -0.06&\cr
   302&225136&  0&    0&   0.96511407& 66.71220273&    3.31& 0.60&     9.9& 1.0&    -6.4& 1.1&    7.86& 0.56  & -8.15& 0.46&  -0.28&\cr
   323&225172&  0&    0&   1.04272795& 49.87020726&    6.15& 0.89&    -8.1& 1.3&   -11.8& 1.3&   -8.01& 0.63  &-10.94& 0.60&  -0.37&\cr
   368&225221&  0&    0&   1.15607578&  2.93722463&    4.83& 1.03&    18.7& 1.1&     0.9& 1.1&   20.68& 1.07  &  1.42& 0.81&   0.98&\cr
   374&225220&  0&    0&   1.16704444& 34.26519797&    6.13& 1.59&     0.0& 0.0&     0.0& 0.0&  -15.80& 1.51  &-30.38& 1.09&   0.00&\cr
   379&225216&  0&    0&   1.17432297& 67.16638761&   10.30& 0.58&    95.1& 1.2&    24.7& 1.3&   95.58& 0.48  & 23.80& 0.45&   0.13&\cr
   399&225276&  0&    0&   1.23302998& 26.64879958&    5.34& 0.77&   109.4& 1.1&   -13.1& 1.1&  112.32& 0.71  &-14.52& 0.47&   1.73&\cr
   466&    58&  0&    0&   1.39547600& 53.17169568&    3.52& 0.83&    16.0& 0.8&    -0.4& 0.8&   16.36& 0.67  & -0.82& 0.57&   0.26&\cr
   472&    71&  0&    0&   1.41549758& 55.70990230&    4.25& 0.78&     7.8& 1.1&    -1.8& 1.2&    8.82& 0.63  & -1.82& 0.68&   0.53&\cr
   496&   100&  0&    0&   1.48628758& 24.56904604&    5.31& 0.73&    34.7& 0.8&   -16.8& 0.8&   34.93& 0.61  &-17.05& 0.49&   0.23&\cr
   502&   111&  0&    0&   1.50224247& 28.27696051&    2.90& 0.77&    -7.0& 1.1&    -4.1& 1.1&   -5.09& 0.63  & -4.69& 0.49&  -0.68&\cr
   504&   112&  0&    0&   1.51336417& 24.91664125&    5.66& 0.82&    82.7& 1.1&   -18.3& 1.0&   85.60& 0.68  &-16.84& 0.52&   1.48&\cr
   525&   145&  0&    0&   1.58393383& 40.89871965&    8.60& 0.73&    83.3& 1.0&     8.2& 1.1&   82.91& 0.74  & 10.65& 0.58&  -0.06&\cr
   598&   236&  0&    0&   1.81025436& 20.55571109&    3.54& 1.00&    11.9& 0.9&     2.1& 0.9&   12.29& 0.95  &  1.16& 0.74&   0.15&\cr
   607&   249&  0&    0&   1.84365756& 26.45088655&    9.37& 0.84&   131.5& 1.2&  -119.4& 1.2&  131.20& 0.83& -117.65& 0.52&  -0.71&\cr
   625&   279&  0&    0&   1.90698995& 40.07851976&    5.37& 0.80&    35.9& 1.4&    -7.8& 1.4&   35.24& 0.71 &  -8.78& 0.55&  -0.19&\cr
   704&   405&  0&    0&   2.19110187& 14.14205698&    5.19& 0.91&    17.7& 0.9&   -33.5& 0.9&   18.87& 0.75 & -33.40& 0.60&   0.30&\cr
   714&   414&  0&    0&   2.20832630& 40.49456869&    3.19& 0.76&     8.4& 1.0&    -6.4& 1.1&    7.09& 0.73 &  -7.99& 0.56&   0.07&\cr
   716&   417&  0&    0&   2.21695394& 25.46285169&    7.78& 0.81&   103.5& 1.1&    55.1& 1.0&  104.61& 0.83 &  54.77& 0.51&   0.47&\cr
   755&   443&  0&    0&   2.31906353& 65.07079864&   17.03& 0.69&   130.9& 1.3&    37.0& 1.4&  128.28& 0.69 &  38.30& 0.62&  -1.01&\cr
   779&     0& 61&    8&   2.40151993& 62.66781255&    4.76& 2.01&     0.0& 0.0&     0.0& 0.0&   -2.50& 1.80 &   2.02& 2.03&   0.00&\cr
   799&   529&  0&    0&   2.46856966& 12.71084240&    5.18& 1.15&    -6.3& 1.3&    -6.0& 1.2&   -6.09& 0.98 &  -5.73& 0.68&  -0.16&\cr
   824&   559&  0&    0&   2.52286733& 25.19788575&    3.13& 0.90&     4.3& 0.9&    10.6& 0.9&    4.16& 0.81 &   9.10& 0.58&  -0.89&\cr
   834&   553&  0&    0&   2.54377402& 64.64672491&    4.74& 0.83&   -29.2& 1.3&    -4.1& 1.4&  -30.17& 0.84 &  -4.08& 0.76&   0.43&\cr
   852&   598&  0&    0&   2.61150772& 28.65292193&    3.65& 0.87&    -6.6& 1.2&   -16.0& 1.1&   -5.04& 0.81 & -16.73& 0.54&   0.09&\cr
   858&   613&  0&    0&   2.63506952& 33.13088570&    4.74& 0.76&   -24.7& 1.2&    -0.6& 1.2&  -25.63& 0.75 &  -0.52& 0.51&   0.48&\cr
   868&   632&  0&    0&   2.66362432&  2.30067277&    2.91& 0.97&     0.1& 1.0&     1.9& 1.1&    0.04& 0.94 &   3.36& 0.67&   0.77&\cr
   871&232121&  0&    0&   2.67528589& 54.89149475&    1.74& 2.12&     0.0& 0.0&     0.0& 0.0&   -1.90& 1.87 &  -0.81& 1.82&   0.00&\cr
   902&   663&  0&    0&   2.78289623& 57.58791462&    4.43& 0.73&   -37.3& 1.0&    11.3& 1.1&  -37.27& 0.53 &  11.00& 0.55&  -0.07&\cr
   909&   685&  0&    0&   2.81532971&  7.94816079&    4.21& 1.02&   -10.8& 0.8&    -8.3& 0.9&  -10.76& 0.85 &  -6.87& 0.65&  -0.53&\cr
   918&   678&  0&    0&   2.84178323& 74.48458019&    6.06& 0.67&    73.3& 0.9&   -13.3& 1.1&   73.95& 0.65 & -15.28& 0.53&   0.62&\cr
   967&   743&  0&    0&   2.99596734& 48.15237121&    5.93& 0.70&    59.5& 1.5&    10.8& 1.5&   60.54& 0.46 &  10.39& 0.45&   0.43&\cr
   969&   756&  0&    0&   3.00550041& 23.47174032&    6.88& 0.76&    87.1& 0.9&     3.8& 0.9&   86.82& 0.66 &   3.73& 0.46&  -0.19&\cr
   972&   754&  0&    0&   3.00807432& 48.68005386&    3.49& 1.01&    17.5& 1.5&    32.4& 1.5&   18.32& 0.78 &  33.73& 0.69&   0.66&\cr
   980&   763&  0&    0&   3.02849022& 47.49116017&    6.68& 0.87&   -13.7& 0.9&   -25.3& 1.1&  -13.46& 0.65 & -24.35& 0.56&  -0.57&\cr
   982&   762&  0&    0&   3.03640453& 48.18225854&    3.46& 0.87&    17.9& 1.1&    -4.5& 1.2&   16.16& 0.60 &  -2.94& 0.56&  -1.11&\cr
   989&   784&  0&    0&   3.06609863& 22.55671394&    2.89& 0.85&    19.8& 0.7&     9.4& 0.7&   19.02& 0.82 &   8.66& 0.53&  -0.73&\cr
\hline
\end{tabular}
\end{table*}

\begin{table*}\addtocounter{table}{-1}
\caption[]{
The right-hand side of the first page of Table A.1, fully available in electronic form at the CDS.  For the sake of clarity, the HIP number 
has been repeated in the leftmost column of the printed version, and the notes have been listed at the bottom of the table.
}
\begin{tabular}{rrrrrrrrrrrrrrrrrrlrrrrrrrrrrrrrrrrrrrrrrrrr}
\hline
 HIP &\multicolumn{2}{c}{$Hp \; \pm$} &   $V_{T2}$ &  \multicolumn{1}{c}{$Hp-V_{T2}$} & $V-I$ & $V-I$ & B &  \multicolumn{2}{c}{$v_r\;\pm$} & 
$M_{Hp}$ &  
 \multicolumn{2}{c}{   $d \;\pm$}&  $A_V$  &      $U$  &    $V$ &    $W$ & g &  P(g) & R.\cr
& & & & & H40 &new & & \multicolumn{2}{c}{(km/s)} & &  \multicolumn{2}{c}{(pc)}& &   \multicolumn{3}{c}{(km/s)}  & & (\%) \cr    
\hline
    21& 7.6923&0.0012& 7.683& 0.009& 1.43& 1.24&*&  -11.72& 0.16& 1.29& 184.4&  28.0&0.07&  -42.5& -31.4&  -1.6&3& 49.7& \cr
    31& 7.7648&0.0020& 7.793&-0.028& 1.46& 1.47&*&    3.52& 0.21&-0.67& 471.6& 131.2&0.07&    6.5&   6.2&  -0.6&5& 61.6& \cr
    36& 7.8503&0.0015& 7.805& 0.045& 1.11& 1.00&0&    0.00& 0.00& 0.00&   0.0&   0.0&0.00&    0.0&   0.0&   0.0&0&   0.& \cr
    73& 7.0364&0.0011& 7.054&-0.018& 1.44& 1.41&*&  -11.91& 0.20&-0.51& 245.6&  35.3&0.60&   24.9&  -1.6&  24.9&3& 90.4& \cr
   119& 6.9834&0.0010& 6.934& 0.049& 1.04& 0.97&*&    0.53& 0.21& 0.95& 146.4&  18.1&0.20&   10.3&   1.9&  -9.6&5& 58.7& \cr
   121& 6.9957&0.0011& 6.964& 0.032& 1.16& 1.09&*&  -15.96& 0.21&-0.33& 272.1&  48.0&0.15&   43.0& -20.2& -21.9&3& 97.7& \cr
   123& 7.3790&0.0013& 7.342& 0.037& 1.13& 1.05&*&  -13.32& 0.22& 0.36& 203.0&  26.3&0.49&   42.0&   7.3&   9.3&3& 90.6& \cr
   136& 6.7635&0.0010& 6.722& 0.041& 1.12& 1.02&*&   -7.69& 0.21& 0.31& 182.6&  25.0&0.15&   -5.6& -24.8& -17.7&3& 77.9& \cr
   142& 7.4726&0.0013& 7.404& 0.069& 0.78& 0.83&*&  -21.68& 0.24& 3.29&  63.7&   2.9&0.16&    6.9& -20.4&   2.1&3& 68.3& \cr
   181& 6.7960&0.0034& 6.980&-0.184& 2.46& 2.30&*&  -21.39& 0.23&-1.98& 377.9&  67.0&0.89&   25.8& -11.4& -17.8&1& 54.4& \cr
   258& 6.7616&0.0008& 6.713& 0.049& 1.08& 0.97&*&  -18.50& 0.21& 0.56& 162.5&  16.3&0.15&  -17.5& -56.3& -23.8&6& 71.4& \cr
   282& 8.0651&0.0018& 8.075&-0.010& 1.39& 1.36&*&  -47.28& 0.22&-0.73& 535.3& 144.5&0.15&   47.3& -73.3& -21.8&2& 44.8& \cr
   288& 8.0688&0.0013& 8.028& 0.041& 1.04& 1.03&*&    3.33& 0.38& 1.36& 197.5&  32.8&0.23&  -70.3& -49.2& -58.6&2& 49.9& \cr
   302& 6.3544&0.0045& 6.520&-0.166& 2.28& 2.22&0&    0.00& 0.00& 0.00&   0.0&   0.0&0.00&    0.0&   0.0&   0.0&0&  0. & \cr
   323& 7.5413&0.0010& 7.475& 0.066& 0.98& 0.85&*&  -16.05& 0.39& 0.92& 192.9&  34.6&0.19&   11.5& -11.8&  -5.6&3& 76.8& \cr
   368& 8.0579&0.0016& 8.045& 0.013& 1.21& 1.21&*&  -16.06& 0.22& 0.83& 270.3&  67.3&0.07&  -19.9& -18.6&   9.9&3& 52.0& \cr
   374& 7.1831&0.0011& 7.255&-0.072& 0.96& 1.73&4&  -23.98& 0.29& 0.27& 216.8&  56.2&0.23&   29.3& -22.9& -13.0&3& 97.4& \cr
   379& 5.8511&0.0006& 5.794& 0.057& 1.02& 0.92&*&  -28.43& 0.20& 0.61&  99.5&   5.7&0.25&  -27.1& -47.2&   1.0&6& 64.3& \cr
   399& 6.3941&0.0010& 6.415&-0.021& 1.34& 1.43&0&    0.00& 0.00& 0.00&   0.0&   0.0&0.00&    0.0&   0.0&   0.0&0&   0.& \cr
   466& 7.3999&0.0009& 7.358& 0.042& 1.14& 1.02&*&   -9.72& 0.22&-0.39& 314.7&  51.6&0.30&  -14.6& -20.3&  -3.3&4& 46.2& \cr
   472& 7.1505&0.0009& 7.115& 0.035& 1.15& 1.06&0&    0.00& 0.00& 0.00&   0.0&   0.0&0.00&    0.0&   0.0&   0.0&0&   0.& \cr
   496& 7.0181&0.0012& 7.067&-0.049& 1.58& 1.60&*&  -14.36& 0.22& 0.37& 199.1&  27.1&0.15&  -18.3& -32.9&  -9.4&3& 57.3& \cr
   502& 6.7361&0.0007& 6.774&-0.038& 1.58& 1.53&*&    9.98& 0.20&-1.14& 350.5&  77.3&0.15&    5.7&   9.9&  -9.0&5& 59.9& \cr
   504& 7.6749&0.0012& 7.608& 0.067& 1.07& 0.85&*&  -10.40& 0.21& 1.15& 188.2&  20.7&0.15&  -54.5& -49.0& -19.4&6& 82.7& \cr
   525& 7.0996&0.0008& 7.062& 0.038& 1.05& 1.05&*&  -73.82& 0.25& 1.49& 121.7&  11.1&0.18&  -14.1& -83.7&  22.6&2& 94.9& \cr
   598& 8.1267&0.0015& 8.126& 0.001& 1.44& 1.29&*&   29.49& 0.23& 0.09& 378.5& 109.4&0.15&  -20.4&  12.5& -20.3&3& 92.4& \cr
   607& 7.4919&0.0011& 7.438& 0.054& 0.99& 0.94&*&   16.44& 0.22& 2.14& 110.4&  10.0&0.14&  -41.5& -47.7& -71.0&2& 90.8& \cr
   625& 7.5541&0.0008& 7.503& 0.051& 0.98& 0.96&*&  -15.02& 0.51& 0.96& 188.0&  20.8&0.23&  -16.3& -29.0&  -6.1&4& 49.3& \cr
   704& 8.0043&0.0013& 7.947& 0.057& 1.09& 0.91&*&  -37.60& 0.22& 1.05& 237.8&  38.2&0.07&   11.2& -56.0&  -0.4&6& 46.4& \cr
   714& 6.9335&0.0012& 6.990&-0.056& 1.59& 1.64&*&  -11.78& 0.54&-1.14& 370.0&  82.0&0.24&    3.8& -19.4&  -8.3&1& 57.5& \cr
   716& 6.4099&0.0006& 6.354& 0.056& 0.97& 0.92&*&   16.11& 0.20& 0.71& 129.3&  12.6&0.15&  -72.5&  -1.4&   6.6&3& 92.4& \cr
   755& 7.1836&0.0010& 7.128& 0.056& 0.93& 0.93&*&    5.00& 0.22& 3.15&  59.7&   2.4&0.15&  -37.0& -14.0&   4.5&3& 63.9& \cr
   779& 9.4783&0.0041& 9.750&-0.272& 2.22& 2.68&9&  -74.31& 0.13&-5.78&3601.3& 771.2&2.47&  158.0& -51.3&  40.7&1& 59.2& *\cr
   799& 8.5304&0.0016& 8.471& 0.059& 1.04& 0.90&*&   -1.34& 0.53& 1.03& 306.6& 100.3&0.07&   11.2&  -2.1&  -3.1&3& 51.6& \cr
   824& 7.8382&0.0022& 7.985&-0.147& 2.15& 2.13&*&   18.91& 0.22&-0.08& 357.1&  87.1&0.15&  -13.7&  19.4&   1.6&3& 86.5& \cr
   834& 8.2793&0.0043& 8.263& 0.016& 1.00& 1.19&1&  -38.93& 0.30& 0.67& 251.4&  47.7&0.61&   46.7& -17.2&  -0.7&3& 98.2& *\cr
   852& 7.0964&0.0048& 7.345&-0.249& 2.64& 2.59&*&  -14.17& 0.34&-0.53& 312.1&  72.7&0.15&   19.4& -16.4& -10.1&3& 86.2& \cr
   858& 6.9629&0.0007& 6.970&-0.007& 1.42& 1.34&*&  -15.05& 0.26&-0.05& 226.9&  37.5&0.24&   28.8&   0.5&  10.9&3& 64.2& \cr
   868& 8.3296&0.0013& 8.298& 0.032& 1.13& 1.09&*&  -40.89& 0.23& 0.10& 428.5& 126.7&0.07&    9.0& -18.4&  36.9&3& 80.2& \cr
   871& 9.1126&0.0393& 9.133&-0.020& 0.73& 1.42&6&   -6.50& 0.30&-1.39& 934.2& 319.8&0.65&   17.1&  -1.4&  -1.3&1& 38.0& *\cr
   902& 6.8469&0.0010& 6.821& 0.026& 1.17& 1.13&*&  -16.90& 0.31&-0.42& 221.4&  30.7&0.54&   40.6&   3.4&  19.2&3& 89.8& \cr
   909& 7.8542&0.0016& 7.806& 0.048& 1.12& 0.98&*&  -24.72& 0.22& 0.29& 315.9&  89.6&0.07&   17.8& -14.4&  15.3&3& 97.3& \cr
   918& 7.3180&0.0012& 7.264& 0.054& 0.98& 0.94&*&    0.32& 0.21& 1.05& 169.9&  18.0&0.12&  -53.6& -24.9& -19.4&3& 61.1& \cr
   967& 6.3113&0.0009& 6.336&-0.025& 1.48& 1.45&*&   14.25& 0.20&-0.07& 174.1&  20.1&0.17&  -51.0&  -9.4&  -2.4&3& 74.4& \cr
   969& 6.7432&0.0006& 6.688& 0.055& 0.97& 0.93&*&    8.57& 0.21& 0.73& 148.5&  15.6&0.15&  -54.1& -21.2& -12.6&3& 56.8& \cr
   972& 8.3410&0.0017& 8.297& 0.044& 1.02& 1.01&0&    0.00& 0.00& 0.00&   0.0&   0.0&0.00&    0.0&   0.0&   0.0&0&   0.& \cr
   980& 7.5299&0.0011& 7.484& 0.046& 1.04& 0.99&*&    4.36& 0.14& 1.31& 162.6&  21.6&0.16&   12.7&   5.4& -18.1&5& 51.5& \cr
   982& 7.7598&0.0015& 7.758& 0.002& 1.28& 1.28&*&  -19.64& 0.21&-0.17& 333.4&  75.6&0.31&  -14.4& -31.5&  -6.3&3& 48.1& \cr
   989& 7.7254&0.0065& 7.919&-0.194& 2.39& 2.35&*&  -10.26& 0.59&-0.17& 354.7&  77.7&0.15&  -33.9& -15.0&  13.4&3& 71.4& \cr
\hline
\end{tabular}

\begin{tabular}{lp{18cm}}
\multicolumn{2}{l}{Notes:}\cr
779 & Very distant star (first-order correction to the differential galactic rotation 
       inaccurate); binary of VV Cep type (M1epIb + B2.6V)\cr 
834& 2002A\&A...395..885D\cr
871& 12.154,1988A\&A...207...37A\cr
\end{tabular}

\end{table*}

\end{document}